\documentclass[twocolumn,ngerman,english,prl,floatfix]{revtex4-2}
\usepackage[T1]{fontenc}
\usepackage[latin9]{inputenc}
\usepackage{amsmath}
\usepackage{graphicx}
\usepackage{babel}
\begin{document}
\title{Splitting of Fermi point of strongly interacting electrons in one dimension:\\ A nonlinear effect of spin-charge separation}
\author{O. Tsyplyatyev}
\affiliation{Institut f{\"u}r Theoretische Physik, Universit{\"a}t Frankfurt,
Max-von-Laue Stra{\ss}e 1, 60438 Frankfurt, Germany}
\date{\today}
\begin{abstract}
A system of one-dimensional electrons interacting via a short-range
potential described by Hubbard model is considered in the regime of strong coupling using the Bethe ansatz approach. We study
its momentum distribution function at zero temperature and find
one additional singularity, at the $3k_{\mathrm{F}}$ point. We identify
that the second singularity is of the same Luttinger liquid type as
the low-energy one at $k_{\mathrm{F}}$. By calculating the spectral
function simultaneously, we show that the second Luttinger liquid
at $3k_{\mathrm{F}}$ is formed by charge modes only, unlike the known
one around $k_{\mathrm{F}}$ consisting of both spin and charge modes.
This result reveals the ability of the spin-charge separation effect to
split the Fermi point of free electrons into two, demonstrating its
robustness beyond the low-energy limit of Luttinger liquid where it
was originally found. 
\end{abstract}
\maketitle
Interactions have a dramatic effect on electrons in one dimension
that has been attracting a significant interest in condensed-matter
physics for a long time \cite{Giamarchi_book}. Their low-energy excitations
become quantised density waves described by the Tomonaga-Luttinger
liquid (TLL) theory based on linearisation of the spectrum around
the Fermi points \cite{Tomonaga50, Luttinger63,Haldane81}. A hallmark
prediction of this theory is separation of the spin and charge degrees
of freedom of the underlying electrons into density waves of two distinct
types with different velocities \cite{Voit93, Schoenhammer92}. Two
linear dispersions originating from the Fermi point were observed
in experiments on magnetotunnelling spectroscopy in semiconductors
\cite{Auslander05,Jompol09}, on photoemission in organic \cite{Claessen02}
and strongly anisotropic \cite{Kim06} crystals, and on time-resolved
microscopy in cold atoms \cite{Vijayan20}, establishing firmly this
phenomenon.
\begin{figure}[b]
\centering

\includegraphics[width=1\columnwidth]{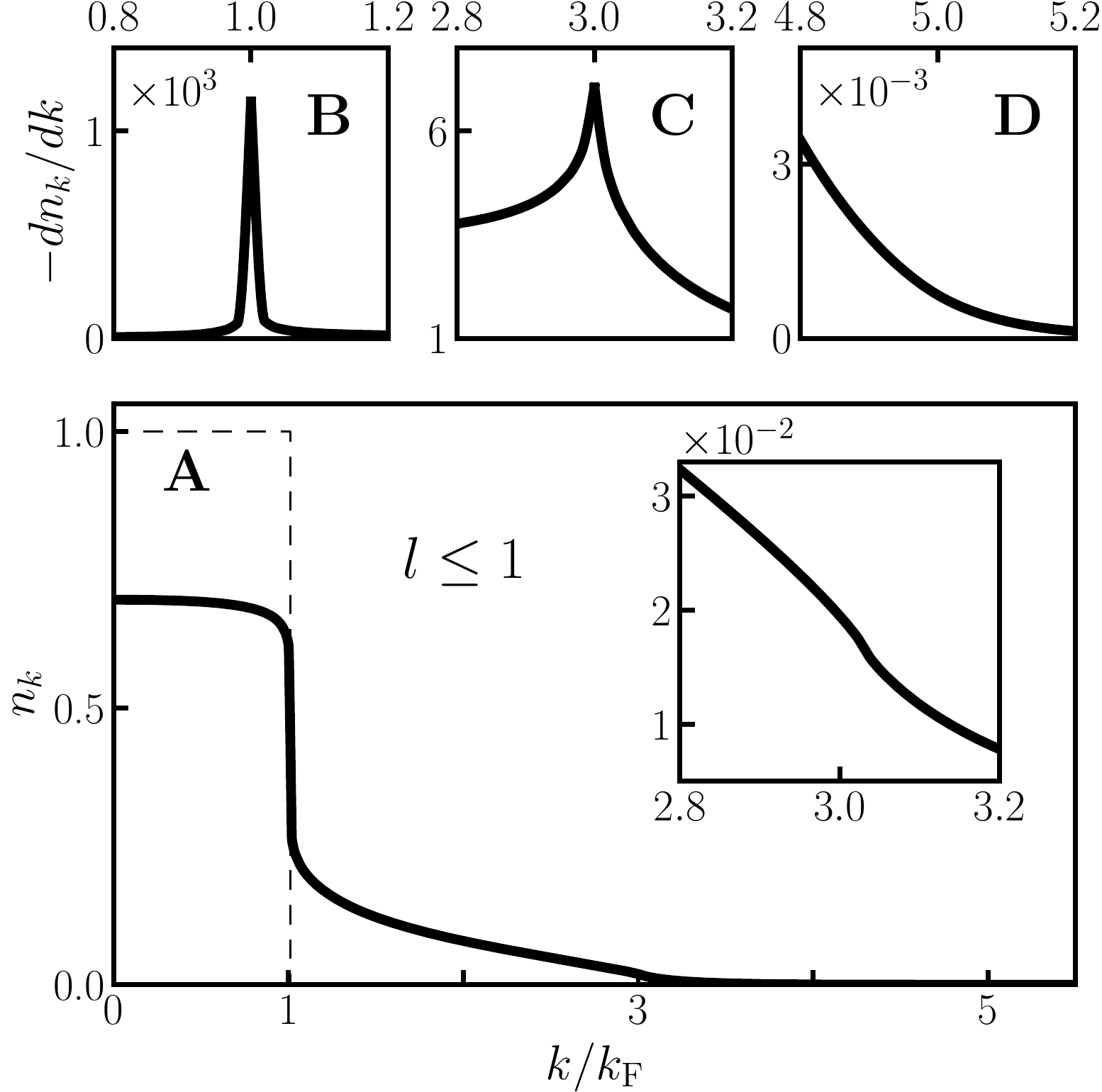}

\caption{\label{fig:Momentum_distribution}(\textbf{A}) Momentum distribution
function $n_{k}$ in the ground state of the model in Eq.~(\ref{eq:Hubbard_model})
evaluated in the $U/t=\infty$ limit using Eqs.~(\ref{eq:ME_charge}-\ref{eq:RMb},\ref{eq:nk}),
where the two leading levels of the hierarchy of modes $l\protect\leq1$
were taken into account in the sum in Eq.~(\ref{eq:nk}), for $N=200$
particles (solid line) and for free particles $U/t=0$ (dashed line).
Inset: Zoom-in around $3k_{\mathrm{F}}$. (\textbf{B}), (\textbf{C}),
and (\textbf{D}) First derivative $dn_{k}/dk$ around the $k_{\mathrm{F}}$,
$3k_{\mathrm{F}}$, and $5k_{\mathrm{F}}$ points respectively.}
\end{figure}

More recently, the theoretical interest was focused on spectral nonlinearity
since it breaks construction of the TLL theory altogether \cite{Samokhin98}
but, on the other hand, is unavoidable at any finite distance from the
Fermi energy in a Fermi system. The Boltzmann equation approach to
weakly interacting Fermi gas predicts a finite relaxation time due
to nonlinearity \cite{Karzig10,Levchenko21,Ristivojevic21}, suggesting
decay of the many-body modes. Application of the mobile impurity model
to Luttinger liquids predicts survival of spin-charge separation at
least in a weak sense, as a singularity--consisting of a mixture
of spin and charge modes--at the spectral edge with nonlinear dispersion
\cite{Imambekov09,Schmidt10,Imambekov12}. At the same time, the continuing
experimental progress is starting to provide information on effects
beyond the low-energy regime \cite{Barak10,Moreno16,Jin19,Wang20,Vianez21}.
In one of these experiments \cite{Vianez21}, the spin-charge separated
modes at low energy were observed to extend to the whole conduction
band forming a pair of parabolic dispersions characterised by two
incommensurate masses, raising the question if the spin-charge separation phenomenon manifests
itself directly in other properties of the whole Fermi sea. 

We explore such a possibility theoretically in this Letter by studying
the momentum distribution function for a Fermi system with short-range
interactions described by the Hubbard model, in which the spin-charge
separation is well-established in the TLL limit \cite{Giamarchi_book}.
Using the microscopic methods of Bethe ansatz in the strong coupling limit ($U=\infty$) not restricted to
low energy \cite{Korepin_book}, we find (at $T=0$) 
one extra divergence at $3k_{\mathrm{F}}$ in addition to the usual
Fermi point at $k_{\mathrm{F}}$, see Fig.~\ref{fig:Momentum_distribution}.
Both singularities are of the same order, in the first derivative $dn_{k}/dk$, revealing the ability of spin-charge separation to split one Fermi point of free electrons into two \cite{FermiSurfaceDefinition}. It is a direct manifestation of this phenomenon in the whole Fermi sea, far away from the TLL limit.
Around the second point, we find a power-law behaviour of $n_{k}$,
$n_{k}\sim\dots+\mathrm{sgn}\left(k-3k_{\mathrm{F}}\right)\left|k-3k_{\mathrm{F}}\right|^{\alpha_{3k_{\mathrm{F}}}}$,
with a real exponent of the TLL type $\alpha_{3k_{\mathrm{F}}}=0.787\pm0.067$.
However, this exponent does not correspond to the known exponents
for spinful or spinless TLLs at $k_{\mathrm{F}}$ \cite{Giamarchi_book}.
By calculating simultaneously the spectral function we show that, out
of the spin-charge separated linear modes around $k_{\mathrm{F}}$,
only the charge branch extends through the nonlinear region to form a TLL around $3k_{\mathrm{F}}$, identifying
it as a TLL of a new kind.

We analyse fermions with spin-1/2 interacting via a short-range potential that are described by the 1D Hubbard model,
\begin{equation}
H=-t\sum_{j,\alpha}\left(c_{j\alpha}^{\dagger}c_{j+1,\alpha}+c_{j\alpha}^{\dagger}c_{j-1,\alpha}\right)+U\sum_{j}n_{j\uparrow}n_{j\downarrow},\label{eq:Hubbard_model}
\end{equation}
where $c_{j\alpha}$ are the Fermi operators at site $j$, $\alpha$
is the spin-1/2 index $\uparrow$ or $\downarrow$, $n_{j\alpha}=c_{j\alpha}^{\dagger}c_{j\alpha}$
is the local density operator of the spin species $\alpha$, $t$
is the hoping amplitude describing the kinetic energy, and $U>0$
is the repulsive two-body interaction energy. Below, we consider the
periodic boundary conditions, $c_{j+L}=c_{j}$, for a 1D lattice consisting
of $L$ sites and for $N$-particle states we  impose
the constraint of low particle density, $N/L\ll1$ \citep{continuum_model}.
In the strong-interaction limit, $U/t=\infty$, the spectrum of the
model in Eq.~(\ref{eq:Hubbard_model}) is given by the following
Lieb-Wu equations \citep{LiebWu68,SM}, 
\begin{align}
Lk_{j} & -P_{s}=2\pi I_{j},\label{eq:LiebWu_charge}\\
Nq_{m} & -2\sum_{l\neq m}^{M}\varphi_{lm}=2\pi J_{m},\label{eq:LiebWu_spin}
\end{align}
where $e^{i2\varphi_{lm}}=-\big(e^{iq_{l}+iq_{m}}+1-2e^{iq_{l}}\big)/\big(e^{iq_{l}+iq_{m}}+1-2e^{iq_{m}}\big)$
are the two-spinon scattering phases, the total spin momentum $P_{s}=\sum_{m}q_{m}$
is defined in the interval of $-\pi..\pi$, and $N$ non-equal integers
$I_{j}$ and $M$ non-equal integers $J_{m}$ define the solution
for the charge $k_{j}$ and spin $q_{m}$ (quasi)momenta of the $N$-particle
state. This solution gives the eigenenergy of the many-body state
as $E=t\sum_{j}k_{j}^{2}/2$ and its momentum as $P=\sum_{j}k_{j}$. 

In the same limit, the eigenstates are factorised, $\left|\Psi\right\rangle =\left|\Psi_{c}\right\rangle \otimes\left|\Psi_{s}\right\rangle $,
into a Slater determinant (like for free particles) for the charge
and a Bethe wave function (like that for a Heisenberg chain) for the
spin degrees of freedom \citep{Ogata90,SM}, 
\begin{align}
\left|\Psi_{c}\right\rangle  & =\frac{1}{L^{N}}\sum_{Q,\mathbf{j}}^{L}\left(-1\right)^{Q}e^{iQ\mathbf{k}\cdot\mathbf{j}}a_{j_{1}}^{\dagger}\cdots a_{j_{N}}^{\dagger}\left|0\right\rangle \label{eq:Psi_c}\\
\left|\Psi_{s}\right\rangle  & =\frac{1}{Z}\sum_{R,x_{1}<\dots<x_{M}}^{N}e^{i\sum_{l<m}\varphi_{R_{l}R_{m}}+iR\mathbf{q}\cdot\mathbf{x}}S_{x_{1}}^{+}\cdots S_{x_{M}}^{+}\left|\Downarrow\right\rangle \label{eq:Psi_s}
\end{align}
where $\mathbf{j}$ are charge coordinates of $N$ particles on the
original Hubbard chain of length $L$, $\mathbf{x}$ are positions
of say $M$ spins pointing up on the spin chain of $N$ spins forming
the spin part of the wave function, and the sums over $Q$ and $R$
run over all possible permutations of $N$ momenta $k_{j}$ and $M$
momenta $q_{m}$ respectively. These wave functions are normalised
to unity, with the non-trivial normalisation factor of the Bethe wave
function being the determinant, $Z^{2}=\det\hat{Q}$, of an $M\times M$
matrix with the following diagonal $Q_{aa}=N-\sum_{l\neq a}^{M}4\left(1-\cos q_{l}\right)/\left(e^{iq_{l}}+e^{-iq_{a}}-2\right)/\left(e^{iq_{a}}+e^{-iq_{l}}-2\right)$
and off-diagonal $Q_{ab}=4\big(1-\cos q_{b}\big)/\big(e^{iq_{a}}+e^{-iq_{b}}-2\big)/\big(e^{iq_{b}}+e^{-iq_{a}}-2\big)$,
with $a\neq b$, elements \citep{Gaudin81}. Here, the spinless Fermi
$a_{j}^{\pm}$ and the purely spin $S_{j}^{\pm}$ operators can be
recombined in the original electron operators $c_{j\alpha}^{\dagger}$
by introducing an insertion(deletion) of the spin-down state at a
given position $x$ in the spin chain operator $I_{x}\left(D_{x}\right)$
as $c_{j\uparrow}^{\dagger}=a_{j}^{\dagger}S_{x}^{+}I_{x}$, $c_{j\downarrow}^{\dagger}=a_{j}^{\dagger}$
$I_{x}$, $c_{j\uparrow}=a_{j}D_{x}S_{x}^{-}$, and $c_{j\downarrow}=a_{j}D_{x}S_{x}^{-}S_{x}^{+}$
\citep{separated_operators}. 

The zero temperature Green function is expressed in terms of the expectation
values of the ladder operators as \citep{Mahan_book} $G_{\alpha}\left(k,E\right)=\sum_{f}\big[\big|\big\langle f\big|c_{k\alpha}^{\dagger}\big|0\big\rangle\big|^{2}/\big(E-E_{f}+i\eta\big)+\big|\big\langle f\big|c_{k\alpha}\big|0\big\rangle\big|^{2}/\big(E+E_{f}-i\eta\big)\big]$,
where $c_{k\alpha}^{\pm}=\sum_{j}c_{j\alpha}^{\pm}e^{\pm ikj}/\sqrt{L}$
is the Fourier transform and $\eta$ is an infinitesimally small real
number. The factorisation of the wave functions makes this calculation
easier since the matrix elements becomes a product of two factors,
$\left\langle f|c_{j\alpha}^{\pm}|0\right\rangle =\left\langle f|c_{j\alpha}^{\pm}|0\right\rangle _{c}\cdot\left\langle f|c_{j\alpha}^{\pm}|0\right\rangle _{s}$,
where $0=\left(\mathbf{k}^{0},\mathbf{q}^{0}\right)$ and $f=\left(\mathbf{k}^{f},\mathbf{q}^{f}\right)$
the momenta of the ground and excited states. The model in Eq.~(\ref{eq:Hubbard_model})
has the symmetry of swapping the spin indices $\uparrow\leftrightarrow\downarrow$,
which its Green function also possesses, $G_{\uparrow}\left(k,E\right)=G_{\downarrow}\left(k,E\right)$.
Therefore, we will consider only $\alpha=\uparrow$. The charge part
of the matrix element is an expectation value with respect to the
state in Eq.~(\ref{eq:Psi_c}) that evaluates as an $N$-fold sum   
over coordinates $\mathbf{j}$ producing a determinant of the Vandermonde
type. Then, application of the generalised Cauchy formula gives the
following result for the annihilation operator $c_{j\uparrow}$ \citep{Penc96,*Penc97p,Penc97}

\begin{multline}
\big<f\big|c_{j\uparrow}\big|0\big>_{c}=\frac{2^{N-1}\sin^{N-1}\left(\frac{P_{s}^{f}-P_{s}^{0}}{2}\right)e^{i\left(P_{0}-P_{f}\right)j}}{L^{N-\frac{1}{2}}}\\
\times\frac{\prod_{i<j}^{N-1}\left(k_{i}^{f}-k_{j}^{f}\right)\prod_{i<j}^{N}\left(k_{i}^{0}-k_{j}^{0}\right)}{\prod_{i,j}^{N,N-1}\left(k_{j}^{f}-k_{i}^{0}\right)},\label{eq:ME_charge}
\end{multline}
where $\big<f\big|c_{j\uparrow}\big|0\big>_{c}\equiv\big<\Psi_{c}^{f}\big|a_{j}\big|\Psi_{c}^{f}\big>$
and the low-density limit, in which $k_{j}^{0,f}\ll1$, is already
taken. 

The spin part part of the matrix elements is an expectation value
with respect to the states in Eq.~(\ref{eq:Psi_s}) that is less
straightforward to evaluate. A mathematical technique for dealing
with these Bethe states analytically was invented in an algebraic
form \citep{Faddeev79}, leading to calculation of the correlation
function for the non-iternarant quantum magnets described by the Heisenberg
model \cite{Kitanine99,*Kitanine00,Caux05}. However, this result
cannot be used here directly since the operators of the Hubbard model
$c_{j\uparrow}^{\pm}$ change the length of the spin chain, making
the constructions of \citep{Faddeev79} for the bra and ket states
incompatible with each other. We resolve this problem by representing
operators of one algebra (for the longer chain) through the other
(for the shorter chain) and the spin operators of the extra site.
Then, explicit evaluation of the expectation value in the spin subspace
of the additional site restores applicability of the methods in \cite{[{See the book by }] [{ and references therein for details.}]Korepin_book},
and we obtain the spin part of the matrix element in terms of the
determinant of an $M\times M$ matrix (see details of this calculation
in \citep{SM}),

\begin{multline}
\big<f\big|c_{j\uparrow}\big|0\big>_{s}=\frac{\det\hat{R}}{Z_{0}Z_{f}}\prod_{i,j}^{M-1,M}\left(e^{iq_{i}^{f}}+e^{-iq_{j}^{0}}-2\right)\\
\times\prod_{i\neq j}^{M-1}\left(e^{iq_{i}^{f}}+e^{-iq_{j}^{f}}-2\right)^{-\frac{1}{2}}\prod_{i\neq j}^{M}\left(e^{iq_{i}^{0}}+e^{-iq_{j}^{0}}-2\right)^{-\frac{1}{2}},\label{eq:ME_spin}
\end{multline}
where $\big<f\big|c_{j\uparrow}\big|0\big>_{s}\equiv\big<\Psi_{s}^{f}\big|D_{x}S^{-}\big|\Psi_{s}^{0}\big>$
and the elements of matrix $\hat{R}$ are 
\begin{align}
R_{ab} & =\frac{e^{iq_{b}^{0}\left(N-1\right)}\prod_{j\neq a}^{M-1}\left(-\frac{e^{iq_{j}^{f}+iq_{b}^{0}}+1-2e^{iq_{j}^{f}}}{e^{iq_{j}^{f}+iq_{b}^{0}}+1-2e^{iq_{b}^{0}}}\right)-1}{\left(e^{-iq_{a}^{f}}-e^{-iq_{b}^{0}}\right)\left(e^{iq_{a}^{f}}+e^{-iq_{b}^{0}}-2\right)},\label{eq:Rab}\\
R_{Mb} & =\frac{e^{ik_{b}^{0}}\prod_{i\neq b}^{M}\left(e^{iq_{i}^{0}}+e^{-iq_{b}^{0}}-2\right)}{\prod_{j}^{M-1}\left(e^{iq_{i}^{f}}+e^{-iq_{b}^{0}}-2\right)}\label{eq:RMb}
\end{align}
for $a<M$ and for $a=M$ respectively. Together Eqs.~(\ref{eq:ME_charge}-\ref{eq:RMb})
give the complete analytical expression for the matrix element for
the 1D Hubbard model. Repeating the same calculation for the matrix
element of the creation operator, $\big\langle f\big|c_{k\uparrow}^{\dagger}\big|0\big\rangle$,
we obtain the same expressions as in Eqs.~(\ref{eq:ME_charge}-\ref{eq:RMb}),
in which the momenta are swapped, $\mathbf{k}^{0},\mathbf{q}^{0}\leftrightarrow\mathbf{k}^{f},\mathbf{q}^{f}$,
and the particle (spin) quantum number is increased by one, $N\rightarrow N+1$
($M\rightarrow M+1$).

\begin{figure}
\centering

\includegraphics[width=1\columnwidth]{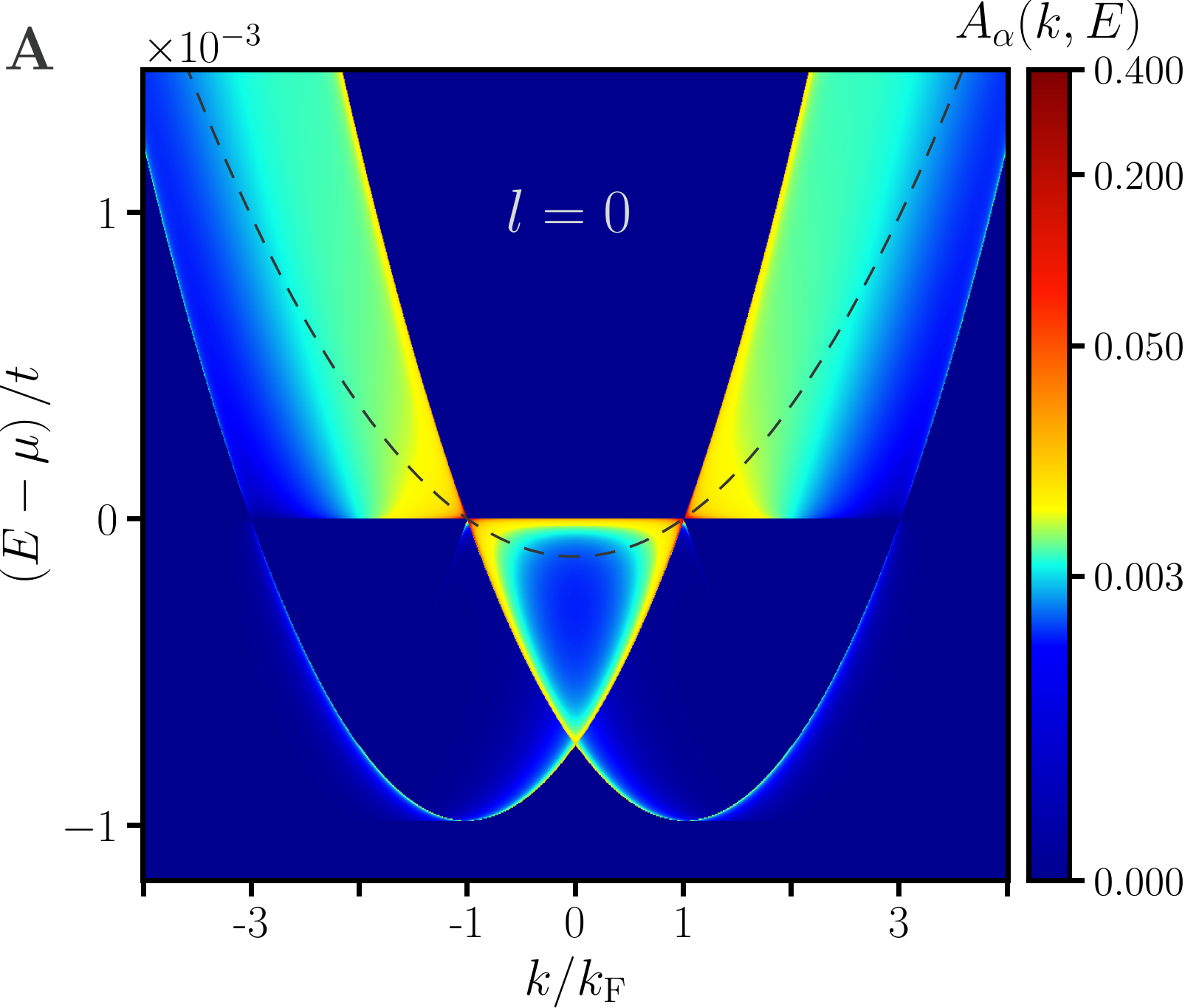}

\includegraphics[width=1\columnwidth]{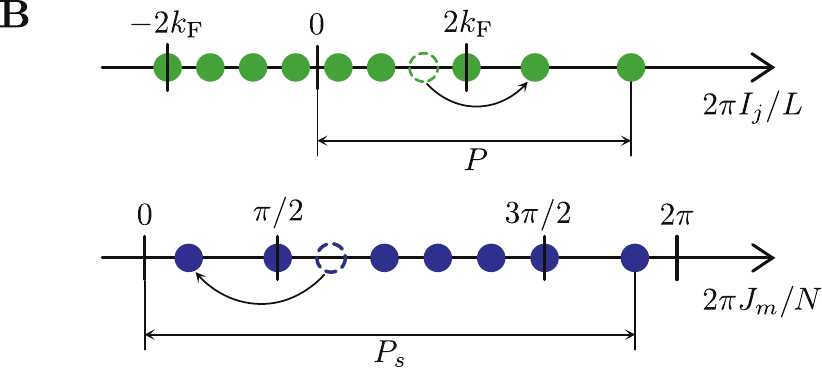}

\caption{\label{fig:Spectral_function}(\textbf{A}) Spectral function $A_{\alpha}\left(k,E\right)$
of the model in Eq.~(\ref{eq:Hubbard_model}) evaluated
in the $U/t=\infty$ limit using Eqs.~(\ref{eq:ME_charge}-\ref{eq:RMb},\ref{eq:A})
for $N=500$ particles, where only the leading level of the hierarchy
$l=0$ was taken into account in the sum in Eq.~(\ref{eq:A}). The
single particle dispersion of free particles $U/t=0$ is superimposed
as dashed line. (\textbf{B}) Two sets of integer numbers, $I_{j}$
for charge and $J_{m}$ for spin degrees of freedom, defining the
Lieb-Wu equations\ (\ref{eq:LiebWu_charge},\ref{eq:LiebWu_spin})
for an excited state with momentum $P$ and the total spin momentum
$P_{s}$; two ``electron-hole'' pairs are shown as examples of lowering
the level of the hierarchy, see text. For a spin-unpolarised system,
the number of spin degrees of freedom is $M=N/2$ that makes the charges'
density twice larger than that of the spins.}
\end{figure}
 The response of a many-body system to a single-particle excitation
at a given momentum and energy is described by the spectral function,
making this observable particularly interesting for the experiments
on spectroscopy. It is related to Green function as $A_{\alpha}\left(k,E\right)=-\textrm{Im}G_{\alpha}\left(k,E\right)\textrm{sgn}\left(E-E_{0}\right)/\pi$
\citep{Mahan_book} giving 

\begin{multline}
A_{\alpha}\left(k,E\right)=\sum_{f}\big|\big<f\big|c_{k\alpha}^{\dagger}\big|0\big>\big|^{2}\delta\left(E-E_{f}+E_{0}\right)\\
+\sum_{f}\big|\big<f\big|c_{k\alpha}\big|0\big>\big|^{2}\delta\left(E+E_{0}-E_{f}\right),\label{eq:A}
\end{multline}
where the sum over the result in Eqs.~(\ref{eq:ME_charge}-\ref{eq:RMb})
\citep{ME_momentum_space} needs to be evaluated over exponentially
many final states $f$. It can be done based on emergence of the hierarchy
of modes: Away from the low-energy regime around the Fermi points
the many-body continuum splits itself into levels (consisting of a
polynomial number of excitations on each of them) according to their
spectral strength, which is proportional to integer powers of a small
parameter $1/L^{2}$ \cite{OT15,*OT16}. 

In the presence of spin and charge degrees of freedom this phenomenon
manifests itself on the microscopic level in the following way. For
the ground states, the charge and spin momenta form two Fermi seas
that correspond to selecting the non-equal integer numbers in Eqs.~(\ref{eq:LiebWu_charge},\ref{eq:LiebWu_spin})
as $I_{j}=-N/2\dots N/2$ and $J_{m}=-\left(N-M\right)/2\dots\left(N+M\right)/2$
\citep{LiebWu68}, see illustration in Fig.~\ref{fig:Spectral_function}B.
The charge part of the spectral amplitude for a generic excitation
above this ground state, $\big|\big<f\big|c_{k\alpha}^{\dagger}\big|0\big>_{c}\big|^{2}$
given by Eq.~(\ref{eq:ME_charge}), is vanishingly small in the
thermodynamic limit since it is proportional to $1/L^{2N}$. However,
the factor $k_{j}^{f}-k_{i}^{0}$ in the denominator of Eq.~(\ref{eq:ME_charge})
produces a singularity that is cut off by $1/L$ cancelling a power
of $1/L^{2}$ in the spectral amplitude each time a charge momentum
of the excited state is equal to a momentum of the ground state. This
property selects a specific set of the excitations, for which a charge
is added above the ground--see the charge state in Fig.~\ref{fig:Spectral_function}B--and
for which the $1/L^{2N}$ factors is canceled altogether making $\big|\big<f\big|c_{k\alpha}^{\dagger}\big|0\big>_{c}\big|^{2}\sim1$.
Adding each ``electron-hole'' pairs of charges on top of these states
multiplies the spectral amplitude by an extra small parameter $1/L^{2}$
since some powers of the normalisation factor $1/L^{2N}$ remain uncanceled. 

For the spin part of the spectral amplitude, $\big|\big<f\big|c_{k\alpha}^{\dagger}\big|0\big>_{s}\big|^{2}$
given by Eq.~(\ref{eq:ME_spin}), emergence of a small parameter
is similar. The normalisation factor $1/\left(Z_{0}Z_{f}\right)$
makes the amplitude proportional to a vanishing in the thermodynamic
limit factor $1/N^{2M}$ and the singularity $1/\big(e^{-iq_{a}^{f}}-e^{-iq_{b}^{0}}\big)$
in the matrix elements in Eq.~(\ref{eq:Rab}) cancels it altogether
for a subset of states, for which only one spin is added on top of
the spin ground states--see the spin state in Fig.~\ref{fig:Spectral_function}B--making
$\big|\big<f\big|c_{k\alpha}^{\dagger}\big|0\big>_{s}\big|^{2}\sim1$.
Adding each ``electron-hole'' pair of spins on top multiplies the
spectral amplitude by an extra small parameter $1/N^{2}$. 

Combined, these two properties result in the hierarchy of modes for
both types of the degrees of freedom, with the spectral power for
the strongest excitations being of the order of $\big|\big<f\big|c_{k\alpha}^{\dagger}\big|0\big>\big|^{2}\sim1$
and the subleading excitations being weak as $\big|\big<f\big|c_{k\alpha}^{\dagger}\big|0\big>\big|^{2}\sim1/\big(N^{2m}L^{2n}\big)$,
where $l=n+m>0$ and $n$ and $m$ are the numbers of extra ``electron-hole''
pairs in the charge and spin Fermi seas respectively. Close to the
Fermi points this hierarchy of modes breaks down. The spectral amplitudes
of all excitations become of the same order forming spin and charge
density waves, and the spinful TLL theory becomes a better approach
for calculating correlation functions \cite{Schoenhammer92,*Voit93}.

Analysing first the nonlinear regime away from the Fermi points, we evaluate the spectral function in Eq.~(\ref{eq:A}) numerically taking into account only the top level of the hierarchy $l=0$ in the sum over $f$ \cite{replicas}. The result is presented  in Fig.~\ref{fig:Spectral_function}A, where the Fermi
momentum is defined by the free particle limit as $k_{\mathrm{F}}=\pi N/2$ and
the corresponding sum rule (which also includes the linear regime around the Fermi points) is already fulfilled as $2\int_{-\infty}^{\infty}dk\int_{-\infty}^{0}dEA\left(k,E\right)/N\simeq61\%$
even for a large number of particles $N=500$. Unlike the case of
spinless fermions \cite{OT15,OT16}, the excitations at the top level
of the hierarchy form a continuum for fermions with spin-1/2 since adding
an electron with spin-1/2 adds simultaneously both charge and spin
with two different momenta $P$ and $P_{s}$, see Fig.~2B. In this continuum, only two
non-equivalent peaks emerge away from the Fermi points: One  connects the $\pm k_{\mathrm{F}}$
points and the other connects the $-k_{\mathrm{F}},3k_{\mathrm{F}}$ points (or equivalently,
the $-3k_{\mathrm{F}},k_{\mathrm{F}}$ points) on the $E=\mu$ line. Around the $\pm k_{\mathrm{F}}$ Fermi points, these nonlinear peaks become two linear peaks that are the manifestation of the collective spin (\emph{i.e} spinon) and charge
(\emph{i.e} holon) modes predicted by the spinful TLL model at low
energy \cite{Schoenhammer92,Voit93}. This identifies
the nonlinear modes as being collective spin and charge excitations as well
(see more details in \citep{SM}), and shows that the spin-charge
separation still manifests itself in observables beyond the linear
TLL limit.

The line shapes of the peaks away from the low-energy region have
the form of divergent power-laws, \emph{e.g.} the particular momentum-dependent
exponent was predicted for the spinon mode (which correspond to the
spectral edge at finite $U$) in \cite{Imambekov09,OT14} and experimentally
confirmed in \citep{Jin19}. The nonlinear holon and spinon modes
were observed directly in the momentum-energy resolved magnetotunneling
experiments in semiconductor quantum wires at intermediate coupling
strength \citep{Moreno16,Vianez21} that strongly suggests their robustness
also for interactions with finite range and of finite strength, beyond
the regime of the present calculation.

The momentum distribution function is another observable of interest
in the many-body systems. It can be obtained from Green function as
$n_{k}=\int dE\allowbreak\Theta\left(E_{0}-E\right)\textrm{Im}G_{\alpha}\left(k,E\right)/\pi$
\citep{Mahan_book} giving
\begin{equation}
n_{k}=\sum_{f}\big|\big<f\big|c_{k\alpha}\big|0\big>\big|^{2}.\label{eq:nk}
\end{equation}
One of the regions of particular interest for this quantity is proximity
of the Fermi point, in which the infinite number the ``electron-hole''
pairs has to be included in the sum over $f$. This can be, at least
partially, accounted for by adding subleading levels of the hierarchy.
The result of the numerical calculation for $N=200$ and $l\leq1$
is presented in Fig.~\ref{fig:Momentum_distribution}A, where the
sum rule already accounts for $2\int dkn_{k}/N\simeq86\%$ of the
particles.

\begin{figure}[t]
	\centering
	\includegraphics[width=1\columnwidth]{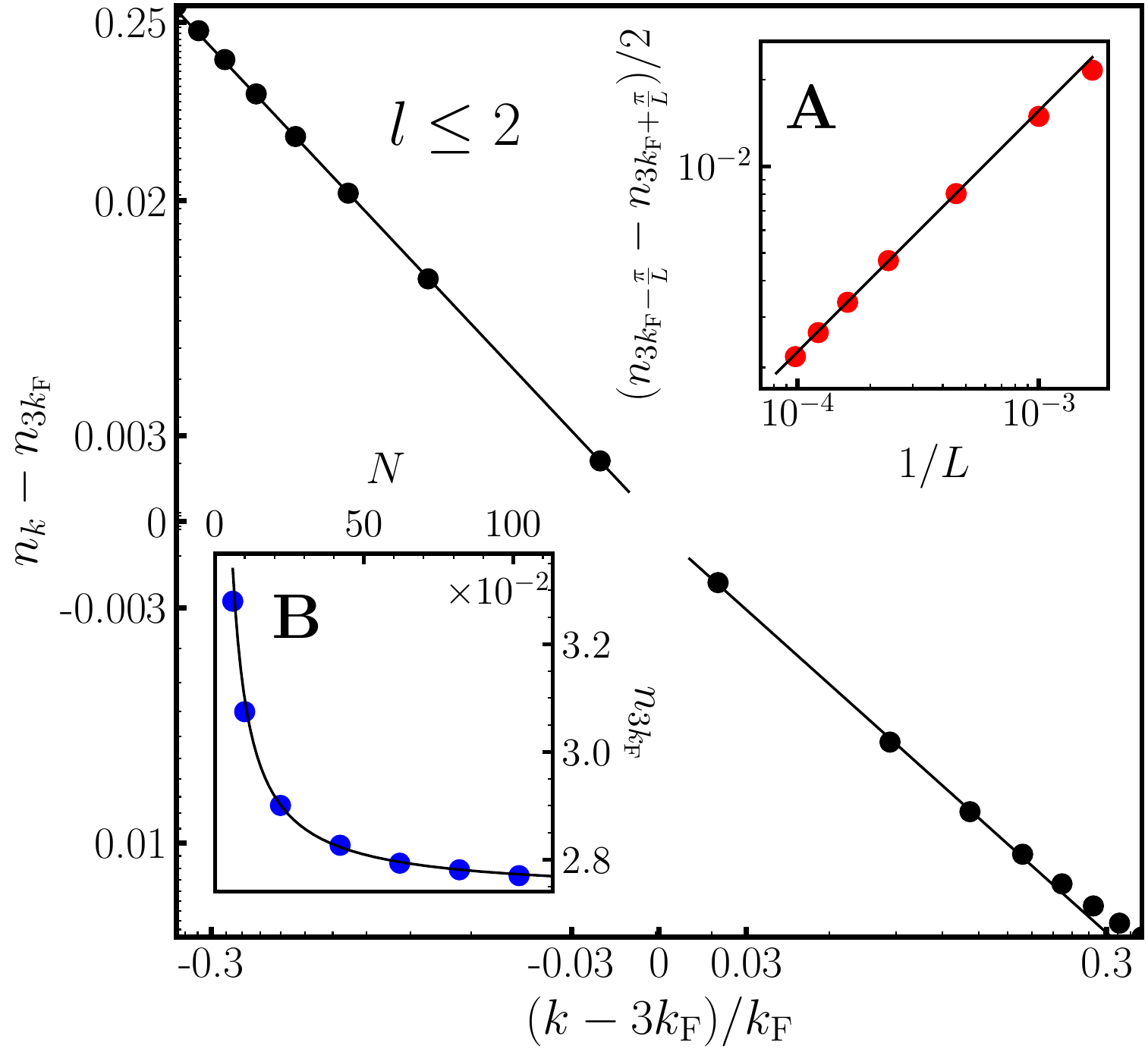}
	\caption{\label{fig:Momentum_distribution_loglog}Momentum distribution function
		$n_{k}$ around $3k_{\mathrm{F}}$ on the log-log scale for $N=80$
		particles, where the three leading levels of the hierarchy of modes
		$l\protect\leq2$ were taken into account in the sum in Eq.~(\ref{eq:nk}).
		The dashed solid lines are power-law functions for $k<3k_{\mathrm{F}}$
		and for $k>3k_{\mathrm{F}}$ giving the exponent as $a_{3k_{\mathrm{F}}}=0.787\pm0.067$,
		where the value is the average of the two and the error bars is the
		difference. Inset A: Finite size cutoff for $n_{k}$ as the function
		of inverse system size $1/L$ on the log-log scale at $3k_{\mathrm{F}}$. The solid line is a power-law fit giving $a_{3k_{\mathrm{F}}}=0.838$,
		within the accuracy of the fitting $n_{k}$ directly. Inset B: The
		value of $n_{k}$ at $3k_{\mathrm{F}}$ as a function of the particle
		number $N$. The solid line is a fit of finite size corrections, $n_{3k_{\mathrm{F}}}=C_{3k_{\mathrm{F}}}+b/N$,
		giving $C_{3k_{\mathrm{F}}}=0.027$.}
\end{figure}
Singularities in $n_{k}$ were introduced as the definition of a Fermi
surface in many-body systems in the Luttinger
theorem \citep{Luttinger60}. Here, we  use this definition to interpret the
result in Fig.~\ref{fig:Momentum_distribution}. Due to the TLL
physics, the singularity at $k_{\mathrm{F}}$ is weaker in 1D \citep{Haldane93},
\emph{e.g.} instead of a discontinuity in $n_{k}$ in $D>1$ dimensions
$n_{k}$ is finite, with a divergence appearing in $dn_{k}/dk$ in
1D, see Fig.~\ref{fig:Momentum_distribution}B. Inspecting the result
in Fig.~\ref{fig:Momentum_distribution}A, we find a second singularity
of same order at $3k_{\mathrm{F}}$, see Fig.~\ref{fig:Momentum_distribution}C.
And we find no divergencies at any other points, \emph{e.g.} see Fig.~\ref{fig:Momentum_distribution}D. The second singularity can be interpreted as a direct manifestation of spin-charge separation beyond the low-energy regime, which facilitates appearance of two Fermi points at different momenta since the density of nonlinear holons twice larger than that of the nonlinear spinons in a spin-unpolarised system.
Note that $n_{k=0}\simeq0.7$ in Fig.~\ref{fig:Momentum_distribution}A
is still smaller than $n_{k=0}=1$ for the free system so $\sim30\%$ of particles is redistributed above $k_{\mathrm{F}}$  providing a significant amplitude around the $3k_{\mathrm{F}}$ point.

The coinciding order of both singularities suggests that the states
around $3k_{\mathrm{F}}$ are also described by a TLL model, which
predicts a power-law behaviour of the momentum distribution function
\citep{Giamarchi_book},
\begin{equation}
n_{k}=C_{(3)k_{\mathrm{F}}}+\dots\mathrm{sgn}\left((3)k_{\mathrm{F}}-k\right)\left|k-(3)k_{\mathrm{F}}\right|^{\alpha_{(3)k_{\mathrm{F}}}},\label{eq:n_k_power_laws}
\end{equation}
where $\alpha_{(3)k_{\mathrm{F}}}$ is a real exponents and $C_{(3)k_{\mathrm{F}}}$
is the value of $n_{k}$ exactly at $\left(3\right)k_{\mathrm{F}}$.
Fitting a linear function on a log-log plot of $n_{k}$ extracts the
exponent around $3k_{\mathrm{F}}$ directly, see Fig.~\ref{fig:Momentum_distribution_loglog}.
Alternatively, the same exponent can be extracted from the finite
size cutoff at $3k_{\mathrm{F}}$, see inset A in Fig.~\ref{fig:Momentum_distribution_loglog}.
Both methods give $a_{3k_{\mathrm{F}}}=0.787\pm0.067$. The scaling
of $n_{3k_{\mathrm{F}}}$ with the system size gives a finite amplitude in the thermodynamic limit, $C_{3k_{\mathrm{F}}}=0.027$,
see inset B in Fig.~\ref{fig:Momentum_distribution_loglog}. In
all of these numerical calculations the number of the levels $l$
used in calculating Eq.~(\ref{eq:nk}) was increased until the next
subleading level was giving only a small correction to the value of
$n_{k}$ at each point. Application of the same procedure around $k_{\textrm{F}}$
gives $C_{k_{\mathrm{F}}}=0.477$ and $\alpha_{k_{\mathrm{F}}}=0.124\pm0.020$
\citep{SM}, which is the well-known result of the TLL theory $\alpha_{k_{\mathrm{F}}}=0.125$
\citep{Schultz90,Frahm90,Kawakami90}. However, only one attempt of
using the TLL approach at $3k_{\mathrm{F}}$ was made in \citep{Penc91},
in which $a_{3k_{\mathrm{F}}}=1.125$ was obtained, suggesting a lower
order (in $d^{2}n_{k}/dk^{2}$) of the second singularity. This discrepancy
can be attributed to using both spinon and holon modes in \citep{Penc91},
while the present microscopic calculation shows in Fig.~\ref{fig:Spectral_function}A
that only the holon modes can form a TLL at $3k_{\mathrm{F}}$.

In conclusions, we have shown that spin-charge separation can split
one Fermi surface into two. Together with the recent experimental
observation of spin-charge splitting of the whole band in \citep{Vianez21},
it demonstrates that this phenomenon is more general than it was originally
anticipated, and also that substantial features of the Fermi gas still
survive in 1D despite formation of a genuine many-body continuum by
the interactions.

I thank Andy Schofield for discussions and helpful comments.
This work was funded by the \foreignlanguage{ngerman}{Deutsche Forschungsgemeinschaft}
(DFG) via project number 461313466.

\bibliographystyle{apsrev4-2}
\bibliography{citations}

\begin{thebibliography}{58}%
\makeatletter
\providecommand \@ifxundefined [1]{%
 \@ifx{#1\undefined}
}%
\providecommand \@ifnum [1]{%
 \ifnum #1\expandafter \@firstoftwo
 \else \expandafter \@secondoftwo
 \fi
}%
\providecommand \@ifx [1]{%
 \ifx #1\expandafter \@firstoftwo
 \else \expandafter \@secondoftwo
 \fi
}%
\providecommand \natexlab [1]{#1}%
\providecommand \enquote  [1]{``#1''}%
\providecommand \bibnamefont  [1]{#1}%
\providecommand \bibfnamefont [1]{#1}%
\providecommand \citenamefont [1]{#1}%
\providecommand \href@noop [0]{\@secondoftwo}%
\providecommand \href [0]{\begingroup \@sanitize@url \@href}%
\providecommand \@href[1]{\@@startlink{#1}\@@href}%
\providecommand \@@href[1]{\endgroup#1\@@endlink}%
\providecommand \@sanitize@url [0]{\catcode `\\12\catcode `\$12\catcode
  `\&12\catcode `\#12\catcode `\^12\catcode `\_12\catcode `\%12\relax}%
\providecommand \@@startlink[1]{}%
\providecommand \@@endlink[0]{}%
\providecommand \url  [0]{\begingroup\@sanitize@url \@url }%
\providecommand \@url [1]{\endgroup\@href {#1}{\urlprefix }}%
\providecommand \urlprefix  [0]{URL }%
\providecommand \Eprint [0]{\href }%
\providecommand \doibase [0]{https://doi.org/}%
\providecommand \selectlanguage [0]{\@gobble}%
\providecommand \bibinfo  [0]{\@secondoftwo}%
\providecommand \bibfield  [0]{\@secondoftwo}%
\providecommand \translation [1]{[#1]}%
\providecommand \BibitemOpen [0]{}%
\providecommand \bibitemStop [0]{}%
\providecommand \bibitemNoStop [0]{.\EOS\space}%
\providecommand \EOS [0]{\spacefactor3000\relax}%
\providecommand \BibitemShut  [1]{\csname bibitem#1\endcsname}%
\let\auto@bib@innerbib\@empty
\bibitem [{\citenamefont {Giamarchi}(2003)}]{Giamarchi_book}%
  \BibitemOpen
  \bibfield  {author} {\bibinfo {author} {\bibfnamefont {T.}~\bibnamefont
  {Giamarchi}},\ }\href@noop {} {\emph {\bibinfo {title} {Quantum physics in
  one dimension}}}\ (\bibinfo  {publisher} {Clarendon press},\ \bibinfo
  {address} {Oxford},\ \bibinfo {year} {2003})\BibitemShut {NoStop}%
\bibitem [{\citenamefont {Tomonaga}(1950)}]{Tomonaga50}%
  \BibitemOpen
  \bibfield  {author} {\bibinfo {author} {\bibfnamefont {S.}~\bibnamefont
  {Tomonaga}},\ }\href@noop {} {\bibfield  {journal} {\bibinfo  {journal}
  {Prog. Theor. Phys.}\ }\textbf {\bibinfo {volume} {5}},\ \bibinfo {pages}
  {544} (\bibinfo {year} {1950})}\BibitemShut {NoStop}%
\bibitem [{\citenamefont {Luttinger}(1963)}]{Luttinger63}%
  \BibitemOpen
  \bibfield  {author} {\bibinfo {author} {\bibfnamefont {J.~M.}\ \bibnamefont
  {Luttinger}},\ }\href@noop {} {\bibfield  {journal} {\bibinfo  {journal} {J.
  Math. Phys.}\ }\textbf {\bibinfo {volume} {4}},\ \bibinfo {pages} {1154}
  (\bibinfo {year} {1963})}\BibitemShut {NoStop}%
\bibitem [{\citenamefont {Haldane}(1981)}]{Haldane81}%
  \BibitemOpen
  \bibfield  {author} {\bibinfo {author} {\bibfnamefont {F.~D.~M.}\
  \bibnamefont {Haldane}},\ }\href@noop {} {\bibfield  {journal} {\bibinfo
  {journal} {J. Phys. C: Solid State Phys.}\ }\textbf {\bibinfo {volume}
  {14}},\ \bibinfo {pages} {2585} (\bibinfo {year} {1981})}\BibitemShut
  {NoStop}%
\bibitem [{\citenamefont {Voit}(1993)}]{Voit93}%
  \BibitemOpen
  \bibfield  {author} {\bibinfo {author} {\bibfnamefont {J.}~\bibnamefont
  {Voit}},\ }\href@noop {} {\bibfield  {journal} {\bibinfo  {journal} {Phys.
  Rev. B}\ }\textbf {\bibinfo {volume} {47}},\ \bibinfo {pages} {6740}
  (\bibinfo {year} {1993})}\BibitemShut {NoStop}%
\bibitem [{\citenamefont {Meden}\ and\ \citenamefont
  {Sch{\"o}nhammer}(1992)}]{Schoenhammer92}%
  \BibitemOpen
  \bibfield  {author} {\bibinfo {author} {\bibfnamefont {V.}~\bibnamefont
  {Meden}}\ and\ \bibinfo {author} {\bibfnamefont {K.}~\bibnamefont
  {Sch{\"o}nhammer}},\ }\href@noop {} {\bibfield  {journal} {\bibinfo
  {journal} {Phys. Rev. B}\ }\textbf {\bibinfo {volume} {46}},\ \bibinfo
  {pages} {15753} (\bibinfo {year} {1992})}\BibitemShut {NoStop}%
\bibitem [{\citenamefont {Auslaender}\ \emph {et~al.}(2005)\citenamefont
  {Auslaender}, \citenamefont {Steinberg}, \citenamefont {Yacoby},
  \citenamefont {Tserkovnyak}, \citenamefont {Halperin}, \citenamefont
  {Baldwin}, \citenamefont {Pfeiffer},\ and\ \citenamefont
  {West}}]{Auslander05}%
  \BibitemOpen
  \bibfield  {author} {\bibinfo {author} {\bibfnamefont {O.~M.}\ \bibnamefont
  {Auslaender}}, \bibinfo {author} {\bibfnamefont {H.}~\bibnamefont
  {Steinberg}}, \bibinfo {author} {\bibfnamefont {A.}~\bibnamefont {Yacoby}},
  \bibinfo {author} {\bibfnamefont {Y.}~\bibnamefont {Tserkovnyak}}, \bibinfo
  {author} {\bibfnamefont {B.~I.}\ \bibnamefont {Halperin}}, \bibinfo {author}
  {\bibfnamefont {K.~W.}\ \bibnamefont {Baldwin}}, \bibinfo {author}
  {\bibfnamefont {L.~N.}\ \bibnamefont {Pfeiffer}},\ and\ \bibinfo {author}
  {\bibfnamefont {K.~W.}\ \bibnamefont {West}},\ }\href@noop {} {\bibfield
  {journal} {\bibinfo  {journal} {Science}\ }\textbf {\bibinfo {volume}
  {308}},\ \bibinfo {pages} {88} (\bibinfo {year} {2005})}\BibitemShut
  {NoStop}%
\bibitem [{\citenamefont {Jompol}\ \emph {et~al.}(2009)\citenamefont {Jompol},
  \citenamefont {Ford}, \citenamefont {Griffiths}, \citenamefont {Farrer},
  \citenamefont {Jones}, \citenamefont {Anderson}, \citenamefont {Ritchie},
  \citenamefont {Silk},\ and\ \citenamefont {Schofield}}]{Jompol09}%
  \BibitemOpen
  \bibfield  {author} {\bibinfo {author} {\bibfnamefont {Y.}~\bibnamefont
  {Jompol}}, \bibinfo {author} {\bibfnamefont {C.~J.~B.}\ \bibnamefont {Ford}},
  \bibinfo {author} {\bibfnamefont {J.~P.}\ \bibnamefont {Griffiths}}, \bibinfo
  {author} {\bibfnamefont {I.}~\bibnamefont {Farrer}}, \bibinfo {author}
  {\bibfnamefont {G.~A.~C.}\ \bibnamefont {Jones}}, \bibinfo {author}
  {\bibfnamefont {D.}~\bibnamefont {Anderson}}, \bibinfo {author}
  {\bibfnamefont {D.~A.}\ \bibnamefont {Ritchie}}, \bibinfo {author}
  {\bibfnamefont {T.~W.}\ \bibnamefont {Silk}},\ and\ \bibinfo {author}
  {\bibfnamefont {A.~J.}\ \bibnamefont {Schofield}},\ }\href@noop {} {\bibfield
   {journal} {\bibinfo  {journal} {Science}\ }\textbf {\bibinfo {volume}
  {325}},\ \bibinfo {pages} {597} (\bibinfo {year} {2009})}\BibitemShut
  {NoStop}%
\bibitem [{\citenamefont {Claessen}\ \emph {et~al.}(2002)\citenamefont
  {Claessen}, \citenamefont {Sing}, \citenamefont {Schwingenschl{\"o}gl},
  \citenamefont {Blaha}, \citenamefont {Dressel},\ and\ \citenamefont
  {Jacobsen}}]{Claessen02}%
  \BibitemOpen
  \bibfield  {author} {\bibinfo {author} {\bibfnamefont {R.}~\bibnamefont
  {Claessen}}, \bibinfo {author} {\bibfnamefont {M.}~\bibnamefont {Sing}},
  \bibinfo {author} {\bibfnamefont {U.}~\bibnamefont {Schwingenschl{\"o}gl}},
  \bibinfo {author} {\bibfnamefont {P.}~\bibnamefont {Blaha}}, \bibinfo
  {author} {\bibfnamefont {M.}~\bibnamefont {Dressel}},\ and\ \bibinfo {author}
  {\bibfnamefont {C.~S.}\ \bibnamefont {Jacobsen}},\ }\href@noop {} {\bibfield
  {journal} {\bibinfo  {journal} {Phys. Rev. Lett.}\ }\textbf {\bibinfo
  {volume} {88}},\ \bibinfo {pages} {096402} (\bibinfo {year}
  {2002})}\BibitemShut {NoStop}%
\bibitem [{\citenamefont {Kim}\ \emph {et~al.}(2006)\citenamefont {Kim},
  \citenamefont {Koh}, \citenamefont {Rotenberg}, \citenamefont {Oh},
  \citenamefont {Eisaki}, \citenamefont {Motoyama}, \citenamefont {Uchida},
  \citenamefont {Tohyama}, \citenamefont {Maekawa}, \citenamefont {Shen},\ and\
  \citenamefont {Kim}}]{Kim06}%
  \BibitemOpen
  \bibfield  {author} {\bibinfo {author} {\bibfnamefont {B.~J.}\ \bibnamefont
  {Kim}}, \bibinfo {author} {\bibfnamefont {H.}~\bibnamefont {Koh}}, \bibinfo
  {author} {\bibfnamefont {E.}~\bibnamefont {Rotenberg}}, \bibinfo {author}
  {\bibfnamefont {S.-J.}\ \bibnamefont {Oh}}, \bibinfo {author} {\bibfnamefont
  {H.}~\bibnamefont {Eisaki}}, \bibinfo {author} {\bibfnamefont
  {N.}~\bibnamefont {Motoyama}}, \bibinfo {author} {\bibfnamefont
  {S.}~\bibnamefont {Uchida}}, \bibinfo {author} {\bibfnamefont
  {T.}~\bibnamefont {Tohyama}}, \bibinfo {author} {\bibfnamefont
  {S.}~\bibnamefont {Maekawa}}, \bibinfo {author} {\bibfnamefont {Z.-X.}\
  \bibnamefont {Shen}},\ and\ \bibinfo {author} {\bibfnamefont
  {C.}~\bibnamefont {Kim}},\ }\href@noop {} {\bibfield  {journal} {\bibinfo
  {journal} {Nat. Phys.}\ }\textbf {\bibinfo {volume} {2}},\ \bibinfo {pages}
  {397} (\bibinfo {year} {2006})}\BibitemShut {NoStop}%
\bibitem [{\citenamefont {Vijayan}\ \emph {et~al.}(2020)\citenamefont
  {Vijayan}, \citenamefont {Sompet}, \citenamefont {Salomon}, \citenamefont
  {Koepsell}, \citenamefont {Hirthe}, \citenamefont {Bohrdt}, \citenamefont
  {Grusdt}, \citenamefont {Bloch},\ and\ \citenamefont {Gross}}]{Vijayan20}%
  \BibitemOpen
  \bibfield  {author} {\bibinfo {author} {\bibfnamefont {J.}~\bibnamefont
  {Vijayan}}, \bibinfo {author} {\bibfnamefont {P.}~\bibnamefont {Sompet}},
  \bibinfo {author} {\bibfnamefont {G.}~\bibnamefont {Salomon}}, \bibinfo
  {author} {\bibfnamefont {J.}~\bibnamefont {Koepsell}}, \bibinfo {author}
  {\bibfnamefont {S.}~\bibnamefont {Hirthe}}, \bibinfo {author} {\bibfnamefont
  {A.}~\bibnamefont {Bohrdt}}, \bibinfo {author} {\bibfnamefont
  {F.}~\bibnamefont {Grusdt}}, \bibinfo {author} {\bibfnamefont
  {I.}~\bibnamefont {Bloch}},\ and\ \bibinfo {author} {\bibfnamefont
  {C.}~\bibnamefont {Gross}},\ }\href@noop {} {\bibfield  {journal} {\bibinfo
  {journal} {Science}\ }\textbf {\bibinfo {volume} {367}},\ \bibinfo {pages}
  {186} (\bibinfo {year} {2020})}\BibitemShut {NoStop}%
\bibitem [{\citenamefont {Samokhin}(1998)}]{Samokhin98}%
  \BibitemOpen
  \bibfield  {author} {\bibinfo {author} {\bibfnamefont {K.~V.}\ \bibnamefont
  {Samokhin}},\ }\href@noop {} {\bibfield  {journal} {\bibinfo  {journal} {J.
  Phys. Condens. Matter}\ }\textbf {\bibinfo {volume} {10}},\ \bibinfo {pages}
  {L533} (\bibinfo {year} {1998})}\BibitemShut {NoStop}%
\bibitem [{\citenamefont {Karzig}\ \emph {et~al.}(2010)\citenamefont {Karzig},
  \citenamefont {Glazman},\ and\ \citenamefont {von Oppen}}]{Karzig10}%
  \BibitemOpen
  \bibfield  {author} {\bibinfo {author} {\bibfnamefont {T.}~\bibnamefont
  {Karzig}}, \bibinfo {author} {\bibfnamefont {L.~I.}\ \bibnamefont
  {Glazman}},\ and\ \bibinfo {author} {\bibfnamefont {F.}~\bibnamefont {von
  Oppen}},\ }\href@noop {} {\bibfield  {journal} {\bibinfo  {journal} {Phys.
  Rev. Lett.}\ }\textbf {\bibinfo {volume} {105}},\ \bibinfo {pages} {226407}
  (\bibinfo {year} {2010})}\BibitemShut {NoStop}%
\bibitem [{\citenamefont {Levchenko}\ and\ \citenamefont
  {Micklitz}(2021)}]{Levchenko21}%
  \BibitemOpen
  \bibfield  {author} {\bibinfo {author} {\bibfnamefont {A.}~\bibnamefont
  {Levchenko}}\ and\ \bibinfo {author} {\bibfnamefont {T.}~\bibnamefont
  {Micklitz}},\ }\href@noop {} {\bibfield  {journal} {\bibinfo  {journal} {J.
  Exp. Theor. Phys.}\ }\textbf {\bibinfo {volume} {132}},\ \bibinfo {pages}
  {675} (\bibinfo {year} {2021})}\BibitemShut {NoStop}%
\bibitem [{\citenamefont {Ristivojevic}\ and\ \citenamefont
  {Matveev}(2021)}]{Ristivojevic21}%
  \BibitemOpen
  \bibfield  {author} {\bibinfo {author} {\bibfnamefont {Z.}~\bibnamefont
  {Ristivojevic}}\ and\ \bibinfo {author} {\bibfnamefont {K.~A.}\ \bibnamefont
  {Matveev}},\ }\href@noop {} {\bibfield  {journal} {\bibinfo  {journal} {Phys.
  Rev. Lett.}\ }\textbf {\bibinfo {volume} {127}},\ \bibinfo {pages} {086803}
  (\bibinfo {year} {2021})}\BibitemShut {NoStop}%
\bibitem [{\citenamefont {Imambekov}\ and\ \citenamefont
  {Glazman}(2009)}]{Imambekov09}%
  \BibitemOpen
  \bibfield  {author} {\bibinfo {author} {\bibfnamefont {A.}~\bibnamefont
  {Imambekov}}\ and\ \bibinfo {author} {\bibfnamefont {L.~I.}\ \bibnamefont
  {Glazman}},\ }\href@noop {} {\bibfield  {journal} {\bibinfo  {journal} {Phys.
  Rev. Lett.}\ }\textbf {\bibinfo {volume} {102}},\ \bibinfo {pages} {126405}
  (\bibinfo {year} {2009})}\BibitemShut {NoStop}%
\bibitem [{\citenamefont {Schmidt}\ \emph {et~al.}(2010)\citenamefont
  {Schmidt}, \citenamefont {Imambekov},\ and\ \citenamefont
  {Glazman}}]{Schmidt10}%
  \BibitemOpen
  \bibfield  {author} {\bibinfo {author} {\bibfnamefont {T.~L.}\ \bibnamefont
  {Schmidt}}, \bibinfo {author} {\bibfnamefont {A.}~\bibnamefont {Imambekov}},\
  and\ \bibinfo {author} {\bibfnamefont {L.~I.}\ \bibnamefont {Glazman}},\
  }\href@noop {} {\bibfield  {journal} {\bibinfo  {journal} {Phys. Rev. Lett.}\
  }\textbf {\bibinfo {volume} {104}},\ \bibinfo {pages} {116403} (\bibinfo
  {year} {2010})}\BibitemShut {NoStop}%
\bibitem [{\citenamefont {Imambekov}\ \emph {et~al.}(2012)\citenamefont
  {Imambekov}, \citenamefont {Schmidt},\ and\ \citenamefont
  {Glazman}}]{Imambekov12}%
  \BibitemOpen
  \bibfield  {author} {\bibinfo {author} {\bibfnamefont {A.}~\bibnamefont
  {Imambekov}}, \bibinfo {author} {\bibfnamefont {T.~L.}\ \bibnamefont
  {Schmidt}},\ and\ \bibinfo {author} {\bibfnamefont {L.~I.}\ \bibnamefont
  {Glazman}},\ }\href@noop {} {\bibfield  {journal} {\bibinfo  {journal} {Rev.
  Mod. Phys.}\ }\textbf {\bibinfo {volume} {84}},\ \bibinfo {pages} {1253}
  (\bibinfo {year} {2012})}\BibitemShut {NoStop}%
\bibitem [{\citenamefont {Barak}\ \emph {et~al.}(2010)\citenamefont {Barak},
  \citenamefont {Steinberg}, \citenamefont {Pfeiffer}, \citenamefont {West},
  \citenamefont {Glazman}, \citenamefont {von Oppen},\ and\ \citenamefont
  {Yacoby}}]{Barak10}%
  \BibitemOpen
  \bibfield  {author} {\bibinfo {author} {\bibfnamefont {G.}~\bibnamefont
  {Barak}}, \bibinfo {author} {\bibfnamefont {H.}~\bibnamefont {Steinberg}},
  \bibinfo {author} {\bibfnamefont {L.~N.}\ \bibnamefont {Pfeiffer}}, \bibinfo
  {author} {\bibfnamefont {K.~W.}\ \bibnamefont {West}}, \bibinfo {author}
  {\bibfnamefont {L.}~\bibnamefont {Glazman}}, \bibinfo {author} {\bibfnamefont
  {F.}~\bibnamefont {von Oppen}},\ and\ \bibinfo {author} {\bibfnamefont
  {A.}~\bibnamefont {Yacoby}},\ }\href@noop {} {\bibfield  {journal} {\bibinfo
  {journal} {Nat. Phys.}\ }\textbf {\bibinfo {volume} {6}},\ \bibinfo {pages}
  {489} (\bibinfo {year} {2010})}\BibitemShut {NoStop}%
\bibitem [{\citenamefont {Moreno}\ \emph {et~al.}(2016)\citenamefont {Moreno},
  \citenamefont {Ford}, \citenamefont {Jin}, \citenamefont {Griffiths},
  \citenamefont {Farrer}, \citenamefont {Jones}, \citenamefont {Ritchie},
  \citenamefont {Tsyplyatyev},\ and\ \citenamefont {Schofield}}]{Moreno16}%
  \BibitemOpen
  \bibfield  {author} {\bibinfo {author} {\bibfnamefont {M.}~\bibnamefont
  {Moreno}}, \bibinfo {author} {\bibfnamefont {C.~J.~B.}\ \bibnamefont {Ford}},
  \bibinfo {author} {\bibfnamefont {Y.}~\bibnamefont {Jin}}, \bibinfo {author}
  {\bibfnamefont {J.~P.}\ \bibnamefont {Griffiths}}, \bibinfo {author}
  {\bibfnamefont {I.}~\bibnamefont {Farrer}}, \bibinfo {author} {\bibfnamefont
  {G.~A.~C.}\ \bibnamefont {Jones}}, \bibinfo {author} {\bibfnamefont {D.~A.}\
  \bibnamefont {Ritchie}}, \bibinfo {author} {\bibfnamefont {O.}~\bibnamefont
  {Tsyplyatyev}},\ and\ \bibinfo {author} {\bibfnamefont {A.~J.}\ \bibnamefont
  {Schofield}},\ }\href@noop {} {\bibfield  {journal} {\bibinfo  {journal}
  {Nat. Commun.}\ }\textbf {\bibinfo {volume} {7}},\ \bibinfo {pages} {12784}
  (\bibinfo {year} {2016})}\BibitemShut {NoStop}%
\bibitem [{\citenamefont {Jin}\ \emph {et~al.}(2019)\citenamefont {Jin},
  \citenamefont {Tsyplyatyev}, \citenamefont {Moreno}, \citenamefont {Anthore},
  \citenamefont {Tan}, \citenamefont {Griffiths}, \citenamefont {Farrer},
  \citenamefont {Ritchie}, \citenamefont {Glazman}, \citenamefont {Schofield},\
  and\ \citenamefont {Ford}}]{Jin19}%
  \BibitemOpen
  \bibfield  {author} {\bibinfo {author} {\bibfnamefont {Y.}~\bibnamefont
  {Jin}}, \bibinfo {author} {\bibfnamefont {O.}~\bibnamefont {Tsyplyatyev}},
  \bibinfo {author} {\bibfnamefont {M.}~\bibnamefont {Moreno}}, \bibinfo
  {author} {\bibfnamefont {A.}~\bibnamefont {Anthore}}, \bibinfo {author}
  {\bibfnamefont {W.}~\bibnamefont {Tan}}, \bibinfo {author} {\bibfnamefont
  {J.}~\bibnamefont {Griffiths}}, \bibinfo {author} {\bibfnamefont
  {I.}~\bibnamefont {Farrer}}, \bibinfo {author} {\bibfnamefont
  {D.}~\bibnamefont {Ritchie}}, \bibinfo {author} {\bibfnamefont
  {L.}~\bibnamefont {Glazman}}, \bibinfo {author} {\bibfnamefont
  {A.}~\bibnamefont {Schofield}},\ and\ \bibinfo {author} {\bibfnamefont
  {C.}~\bibnamefont {Ford}},\ }\href@noop {} {\bibfield  {journal} {\bibinfo
  {journal} {Nat. Commun.}\ }\textbf {\bibinfo {volume} {10}},\ \bibinfo
  {pages} {2821} (\bibinfo {year} {2019})}\BibitemShut {NoStop}%
\bibitem [{\citenamefont {Wang}\ \emph {et~al.}(2020)\citenamefont {Wang},
  \citenamefont {Zhao}, \citenamefont {Shi}, \citenamefont {Wu}, \citenamefont
  {Zhao}, \citenamefont {Jiang}, \citenamefont {Watanabe}, \citenamefont
  {Taniguchi}, \citenamefont {Zettl}, \citenamefont {Zhou},\ and\ \citenamefont
  {Wang}}]{Wang20}%
  \BibitemOpen
  \bibfield  {author} {\bibinfo {author} {\bibfnamefont {S.}~\bibnamefont
  {Wang}}, \bibinfo {author} {\bibfnamefont {S.}~\bibnamefont {Zhao}}, \bibinfo
  {author} {\bibfnamefont {Z.}~\bibnamefont {Shi}}, \bibinfo {author}
  {\bibfnamefont {F.}~\bibnamefont {Wu}}, \bibinfo {author} {\bibfnamefont
  {Z.}~\bibnamefont {Zhao}}, \bibinfo {author} {\bibfnamefont {L.}~\bibnamefont
  {Jiang}}, \bibinfo {author} {\bibfnamefont {K.}~\bibnamefont {Watanabe}},
  \bibinfo {author} {\bibfnamefont {T.}~\bibnamefont {Taniguchi}}, \bibinfo
  {author} {\bibfnamefont {A.}~\bibnamefont {Zettl}}, \bibinfo {author}
  {\bibfnamefont {C.}~\bibnamefont {Zhou}},\ and\ \bibinfo {author}
  {\bibfnamefont {F.}~\bibnamefont {Wang}},\ }\href@noop {} {\bibfield
  {journal} {\bibinfo  {journal} {Nat. Mater.}\ }\textbf {\bibinfo {volume}
  {19}},\ \bibinfo {pages} {986} (\bibinfo {year} {2020})}\BibitemShut
  {NoStop}%
\bibitem [{\citenamefont {Vianez}\ \emph {et~al.}(2021)\citenamefont {Vianez},
  \citenamefont {Jin}, \citenamefont {Moreno}, \citenamefont {Anirban},
  \citenamefont {Anthore}, \citenamefont {Tan}, \citenamefont {Griffiths},
  \citenamefont {Farrer}, \citenamefont {Ritchie}, \citenamefont {Schofield},
  \citenamefont {Tsyplyatyev},\ and\ \citenamefont {Ford}}]{Vianez21}%
  \BibitemOpen
  \bibfield  {author} {\bibinfo {author} {\bibfnamefont {P.~M.~T.}\
  \bibnamefont {Vianez}}, \bibinfo {author} {\bibfnamefont {Y.}~\bibnamefont
  {Jin}}, \bibinfo {author} {\bibfnamefont {M.}~\bibnamefont {Moreno}},
  \bibinfo {author} {\bibfnamefont {A.~S.}\ \bibnamefont {Anirban}}, \bibinfo
  {author} {\bibfnamefont {A.}~\bibnamefont {Anthore}}, \bibinfo {author}
  {\bibfnamefont {W.~K.}\ \bibnamefont {Tan}}, \bibinfo {author} {\bibfnamefont
  {J.~P.}\ \bibnamefont {Griffiths}}, \bibinfo {author} {\bibfnamefont
  {I.}~\bibnamefont {Farrer}}, \bibinfo {author} {\bibfnamefont {D.~A.}\
  \bibnamefont {Ritchie}}, \bibinfo {author} {\bibfnamefont {A.~J.}\
  \bibnamefont {Schofield}}, \bibinfo {author} {\bibfnamefont {O.}~\bibnamefont
  {Tsyplyatyev}},\ and\ \bibinfo {author} {\bibfnamefont {C.~J.~B.}\
  \bibnamefont {Ford}},\ }\href@noop {} {\bibfield  {journal} {\bibinfo
  {journal} {arXiv:2102.05584}\ } (\bibinfo {year} {2021})}\BibitemShut
  {NoStop}%
\bibitem [{\citenamefont {Korepin}\ \emph {et~al.}(1993)\citenamefont
  {Korepin}, \citenamefont {Bogoliubov},\ and\ \citenamefont
  {Izergin}}]{Korepin_book}%
  \BibitemOpen
  \bibfield  {author} {\bibinfo {author} {\bibfnamefont {V.~E.}\ \bibnamefont
  {Korepin}}, \bibinfo {author} {\bibfnamefont {N.~M.}\ \bibnamefont
  {Bogoliubov}},\ and\ \bibinfo {author} {\bibfnamefont {A.~G.}\ \bibnamefont
  {Izergin}},\ }\href@noop {} {\emph {\bibinfo {title} {Quantum inverse
  scattering methods and correlation functions}}}\ (\bibinfo  {publisher}
  {Cambridge University Press},\ \bibinfo {year} {1993})\BibitemShut {NoStop}%
\bibitem [{Fer()}]{FermiSurfaceDefinition}%
  \BibitemOpen
  \href@noop {} {}\bibinfo {note} {The definition of a Fermi point (or surface)
  for a many-body system as a special point in $n_{k}$ was given by Luttinger
  and Haldane \cite{Luttinger60,Haldane93}.}\BibitemShut {Stop}%
\bibitem [{con()}]{continuum_model}%
  \BibitemOpen
  \href@noop {} {}\bibinfo {note} {This limit corresponds to the continuum
  model with a finite mass and a $\delta$-functional density-density
  interaction, in which the mass $m$ is inversely proportional to the hopping
  amplitude $t$, the interaction strength is proportional to $U$, and the
  lattice parameter only plays the role of an ultraviolet cutoff.}\BibitemShut
  {Stop}%
\bibitem [{\citenamefont {Lieb}\ and\ \citenamefont {Wu}(1968)}]{LiebWu68}%
  \BibitemOpen
  \bibfield  {author} {\bibinfo {author} {\bibfnamefont {E.~H.}\ \bibnamefont
  {Lieb}}\ and\ \bibinfo {author} {\bibfnamefont {F.~Y.}\ \bibnamefont {Wu}},\
  }\href@noop {} {\bibfield  {journal} {\bibinfo  {journal} {Phys. Rev. Lett.}\
  }\textbf {\bibinfo {volume} {20}},\ \bibinfo {pages} {1445} (\bibinfo {year}
  {1968})}\BibitemShut {NoStop}%
\bibitem [{SM()}]{SM}%
  \BibitemOpen
  \href@noop {} {}\bibinfo {note} {See Supplementary Material at \url{https://}
  for details, which includes
  Refs.~\onlinecite{Bethe31,Gaudin67,Yang67,Gaudin_book,Essler_book,Slavnov89,Korepin82,Drinfeld83}.}\BibitemShut
  {Stop}%
\bibitem [{\citenamefont {Ogata}\ and\ \citenamefont {Shiba}(1990)}]{Ogata90}%
  \BibitemOpen
  \bibfield  {author} {\bibinfo {author} {\bibfnamefont {M.}~\bibnamefont
  {Ogata}}\ and\ \bibinfo {author} {\bibfnamefont {H.}~\bibnamefont {Shiba}},\
  }\href@noop {} {\bibfield  {journal} {\bibinfo  {journal} {Phys. Rev. B}\
  }\textbf {\bibinfo {volume} {41}},\ \bibinfo {pages} {2326} (\bibinfo {year}
  {1990})}\BibitemShut {NoStop}%
\bibitem [{\citenamefont {Gaudin}\ \emph {et~al.}(1981)\citenamefont {Gaudin},
  \citenamefont {McCoy},\ and\ \citenamefont {Wu}}]{Gaudin81}%
  \BibitemOpen
  \bibfield  {author} {\bibinfo {author} {\bibfnamefont {M.}~\bibnamefont
  {Gaudin}}, \bibinfo {author} {\bibfnamefont {B.~M.}\ \bibnamefont {McCoy}},\
  and\ \bibinfo {author} {\bibfnamefont {T.~T.}\ \bibnamefont {Wu}},\
  }\href@noop {} {\bibfield  {journal} {\bibinfo  {journal} {Phys. Rev. D}\
  }\textbf {\bibinfo {volume} {23}},\ \bibinfo {pages} {417} (\bibinfo {year}
  {1981})}\BibitemShut {NoStop}%
\bibitem [{sep()}]{separated_operators}%
  \BibitemOpen
  \href@noop {} {}\bibinfo {note} {While the $a^\pm_j$ and $S^\pm_x$ operators
  obey the regular Fermi and spin commutation rules, the insertion(deletion)
  operator $I_x$($D_x$) does not. In general, they also do not commute with the
  spin operators.}\BibitemShut {Stop}%
\bibitem [{\citenamefont {Mahan}(1990)}]{Mahan_book}%
  \BibitemOpen
  \bibfield  {author} {\bibinfo {author} {\bibfnamefont {G.~D.}\ \bibnamefont
  {Mahan}},\ }\href@noop {} {\emph {\bibinfo {title} {Many-particle physics}}}\
  (\bibinfo  {publisher} {Plenum Press},\ \bibinfo {year} {1990})\BibitemShut
  {NoStop}%
\bibitem [{\citenamefont {Penc}\ \emph {et~al.}(1996)\citenamefont {Penc},
  \citenamefont {Hallberg}, \citenamefont {Mila},\ and\ \citenamefont
  {Shiba}}]{Penc96}%
  \BibitemOpen
  \bibfield  {author} {\bibinfo {author} {\bibfnamefont {K.}~\bibnamefont
  {Penc}}, \bibinfo {author} {\bibfnamefont {K.}~\bibnamefont {Hallberg}},
  \bibinfo {author} {\bibfnamefont {F.}~\bibnamefont {Mila}},\ and\ \bibinfo
  {author} {\bibfnamefont {H.}~\bibnamefont {Shiba}},\ }\href@noop {}
  {\bibfield  {journal} {\bibinfo  {journal} {Phys. Rev. Lett.}\ }\textbf
  {\bibinfo {volume} {77}},\ \bibinfo {pages} {1390} (\bibinfo {year}
  {1996})}\BibitemShut {NoStop}%
\bibitem [{\citenamefont {Penc}\ \emph {et~al.}(1997)\citenamefont {Penc},
  \citenamefont {Hallberg}, \citenamefont {Mila},\ and\ \citenamefont
  {Shiba}}]{Penc97p}%
  \BibitemOpen
  \bibfield  {author} {\bibinfo {author} {\bibfnamefont {K.}~\bibnamefont
  {Penc}}, \bibinfo {author} {\bibfnamefont {K.}~\bibnamefont {Hallberg}},
  \bibinfo {author} {\bibfnamefont {F.}~\bibnamefont {Mila}},\ and\ \bibinfo
  {author} {\bibfnamefont {H.}~\bibnamefont {Shiba}},\ }\href@noop {}
  {\bibfield  {journal} {\bibinfo  {journal} {Phys. Rev. B}\ }\textbf {\bibinfo
  {volume} {55}},\ \bibinfo {pages} {15475} (\bibinfo {year}
  {1997})}\BibitemShut {NoStop}%
\bibitem [{\citenamefont {Penc}\ and\ \citenamefont {Serhan}(1997)}]{Penc97}%
  \BibitemOpen
  \bibfield  {author} {\bibinfo {author} {\bibfnamefont {K.}~\bibnamefont
  {Penc}}\ and\ \bibinfo {author} {\bibfnamefont {M.}~\bibnamefont {Serhan}},\
  }\href@noop {} {\bibfield  {journal} {\bibinfo  {journal} {Phys. Rev. B}\
  }\textbf {\bibinfo {volume} {46}},\ \bibinfo {pages} {6555} (\bibinfo {year}
  {1997})}\BibitemShut {NoStop}%
\bibitem [{\citenamefont {Sklyanin}\ \emph {et~al.}(1979)\citenamefont
  {Sklyanin}, \citenamefont {Takhtajan},\ and\ \citenamefont
  {Faddeev}}]{Faddeev79}%
  \BibitemOpen
  \bibfield  {author} {\bibinfo {author} {\bibfnamefont {E.~K.}\ \bibnamefont
  {Sklyanin}}, \bibinfo {author} {\bibfnamefont {L.~A.}\ \bibnamefont
  {Takhtajan}},\ and\ \bibinfo {author} {\bibfnamefont {L.~D.}\ \bibnamefont
  {Faddeev}},\ }\href@noop {} {\bibfield  {journal} {\bibinfo  {journal}
  {Theor. Math. Phys.}\ }\textbf {\bibinfo {volume} {40}},\ \bibinfo {pages}
  {688} (\bibinfo {year} {1979})}\BibitemShut {NoStop}%
\bibitem [{\citenamefont {Kitanine}\ \emph {et~al.}(1999)\citenamefont
  {Kitanine}, \citenamefont {Maillet},\ and\ \citenamefont
  {Tetras}}]{Kitanine99}%
  \BibitemOpen
  \bibfield  {author} {\bibinfo {author} {\bibfnamefont {N.}~\bibnamefont
  {Kitanine}}, \bibinfo {author} {\bibfnamefont {J.}~\bibnamefont {Maillet}},\
  and\ \bibinfo {author} {\bibfnamefont {V.}~\bibnamefont {Tetras}},\
  }\href@noop {} {\bibfield  {journal} {\bibinfo  {journal} {Nucl. Phys. B}\
  }\textbf {\bibinfo {volume} {554}},\ \bibinfo {pages} {647} (\bibinfo {year}
  {1999})}\BibitemShut {NoStop}%
\bibitem [{\citenamefont {Kitanine}\ and\ \citenamefont
  {Maillet}(2000)}]{Kitanine00}%
  \BibitemOpen
  \bibfield  {author} {\bibinfo {author} {\bibfnamefont {N.}~\bibnamefont
  {Kitanine}}\ and\ \bibinfo {author} {\bibfnamefont {J.~M.}\ \bibnamefont
  {Maillet}},\ }\href@noop {} {\bibfield  {journal} {\bibinfo  {journal} {Nucl.
  Phys. B}\ }\textbf {\bibinfo {volume} {567}} (\bibinfo {year}
  {2000})}\BibitemShut {NoStop}%
\bibitem [{\citenamefont {Caux}\ and\ \citenamefont {Maillet}(2005)}]{Caux05}%
  \BibitemOpen
  \bibfield  {author} {\bibinfo {author} {\bibfnamefont {J.-S.}\ \bibnamefont
  {Caux}}\ and\ \bibinfo {author} {\bibfnamefont {J.~M.}\ \bibnamefont
  {Maillet}},\ }\href@noop {} {\bibfield  {journal} {\bibinfo  {journal} {Phys.
  Rev. Lett.}\ }\textbf {\bibinfo {volume} {95}},\ \bibinfo {pages} {077201}
  (\bibinfo {year} {2005})}\BibitemShut {NoStop}%
\bibitem [{ME_()}]{ME_momentum_space}%
  \BibitemOpen
  \href@noop {} {}\bibinfo {note} {Owning to the translational invariance of
  the model in Eq. (\ref{eq:Hubbard_model}) the expectation values of the
  $c^\pm_k$ operator is related to the result in Eqs.\
  (\ref{eq:ME_charge}-\ref{eq:RMb}) in a trivial way,
  $\big<f\big|c_{k\alpha}^\pm\big|0\big>=\sqrt{L}\big<f\big|c_{1\alpha}^\pm\big|0\big>\delta\big(k\mp
  P_{f}\big)$.}\BibitemShut {Stop}%
\bibitem [{\citenamefont {Tsyplyatyev}\ \emph {et~al.}(2015)\citenamefont
  {Tsyplyatyev}, \citenamefont {Schofield}, \citenamefont {Jin}, \citenamefont
  {Moreno}, \citenamefont {Tan}, \citenamefont {Ford}, \citenamefont
  {Griffiths}, \citenamefont {Farrer}, \citenamefont {Jones},\ and\
  \citenamefont {Ritchie}}]{OT15}%
  \BibitemOpen
  \bibfield  {author} {\bibinfo {author} {\bibfnamefont {O.}~\bibnamefont
  {Tsyplyatyev}}, \bibinfo {author} {\bibfnamefont {A.~J.}\ \bibnamefont
  {Schofield}}, \bibinfo {author} {\bibfnamefont {Y.}~\bibnamefont {Jin}},
  \bibinfo {author} {\bibfnamefont {M.}~\bibnamefont {Moreno}}, \bibinfo
  {author} {\bibfnamefont {W.~K.}\ \bibnamefont {Tan}}, \bibinfo {author}
  {\bibfnamefont {C.~J.~B.}\ \bibnamefont {Ford}}, \bibinfo {author}
  {\bibfnamefont {J.~P.}\ \bibnamefont {Griffiths}}, \bibinfo {author}
  {\bibfnamefont {I.}~\bibnamefont {Farrer}}, \bibinfo {author} {\bibfnamefont
  {G.~A.~C.}\ \bibnamefont {Jones}},\ and\ \bibinfo {author} {\bibfnamefont
  {D.~A.}\ \bibnamefont {Ritchie}},\ }\href@noop {} {\bibfield  {journal}
  {\bibinfo  {journal} {Phys. Rev. Lett.}\ }\textbf {\bibinfo {volume} {114}},\
  \bibinfo {pages} {196401} (\bibinfo {year} {2015})}\BibitemShut {NoStop}%
\bibitem [{\citenamefont {Tsyplyatyev}\ \emph {et~al.}(2016)\citenamefont
  {Tsyplyatyev}, \citenamefont {Schofield}, \citenamefont {Jin}, \citenamefont
  {Moreno}, \citenamefont {Tan}, \citenamefont {Ford}, \citenamefont
  {Griffiths}, \citenamefont {Farrer}, \citenamefont {Jones},\ and\
  \citenamefont {Ritchie}}]{OT16}%
  \BibitemOpen
  \bibfield  {author} {\bibinfo {author} {\bibfnamefont {O.}~\bibnamefont
  {Tsyplyatyev}}, \bibinfo {author} {\bibfnamefont {A.~J.}\ \bibnamefont
  {Schofield}}, \bibinfo {author} {\bibfnamefont {Y.}~\bibnamefont {Jin}},
  \bibinfo {author} {\bibfnamefont {M.}~\bibnamefont {Moreno}}, \bibinfo
  {author} {\bibfnamefont {W.~K.}\ \bibnamefont {Tan}}, \bibinfo {author}
  {\bibfnamefont {C.~J.~B.}\ \bibnamefont {Ford}}, \bibinfo {author}
  {\bibfnamefont {J.~P.}\ \bibnamefont {Griffiths}}, \bibinfo {author}
  {\bibfnamefont {I.}~\bibnamefont {Farrer}}, \bibinfo {author} {\bibfnamefont
  {G.~A.~C.}\ \bibnamefont {Jones}},\ and\ \bibinfo {author} {\bibfnamefont
  {D.~A.}\ \bibnamefont {Ritchie}},\ }\href@noop {} {\bibfield  {journal}
  {\bibinfo  {journal} {Phys. Rev. B}\ }\textbf {\bibinfo {volume} {93}},\
  \bibinfo {pages} {075147} (\bibinfo {year} {2016})}\BibitemShut {NoStop}%
\bibitem [{rep()}]{replicas}%
  \BibitemOpen
  \href@noop {} {}\bibinfo {note} {Addition of extra excitations with $l>0$ to
  the spectral function in Eq.\ (\ref{eq:A}) produces extra levels of the
  hierarchy beyond the low-energy limit that correspond to mirroring with
  respect to the $E=\mu$ line and translating by integer multiples of
  $2k_\mathrm{F}$ and $4k_\mathrm{F}$ the continuum picture of the top level
  $l=0$ in Fig.\ \ref{fig:Spectral_function}. However, the overall amplitude of
  these replicas is proportional to positive integer powers of the small
  parameters $1/L^2$ and $1/N^2$.}\BibitemShut {Stop}%
\bibitem [{\citenamefont {Tsyplyatyev}\ and\ \citenamefont
  {Schofield}(2014)}]{OT14}%
  \BibitemOpen
  \bibfield  {author} {\bibinfo {author} {\bibfnamefont {O.}~\bibnamefont
  {Tsyplyatyev}}\ and\ \bibinfo {author} {\bibfnamefont {A.~J.}\ \bibnamefont
  {Schofield}},\ }\href@noop {} {\bibfield  {journal} {\bibinfo  {journal}
  {Phys. Rev. B}\ }\textbf {\bibinfo {volume} {90}},\ \bibinfo {pages} {014309}
  (\bibinfo {year} {2014})}\BibitemShut {NoStop}%
\bibitem [{\citenamefont {Luttinger}(1960)}]{Luttinger60}%
  \BibitemOpen
  \bibfield  {author} {\bibinfo {author} {\bibfnamefont {J.~M.}\ \bibnamefont
  {Luttinger}},\ }\href@noop {} {\bibfield  {journal} {\bibinfo  {journal}
  {Phys. Rev.}\ }\textbf {\bibinfo {volume} {119}},\ \bibinfo {pages} {1153}
  (\bibinfo {year} {1960})}\BibitemShut {NoStop}%
\bibitem [{\citenamefont {Haldane}(1994)}]{Haldane93}%
  \BibitemOpen
  \bibfield  {author} {\bibinfo {author} {\bibfnamefont {F.~D.~M.}\
  \bibnamefont {Haldane}},\ }in\ \href@noop {} {\emph {\bibinfo {booktitle}
  {Proceedings of the International School of Physics ``Enrico Fermi'', Course
  CXXI: ``Perspectives in Many-Particle Physics''}}},\ \bibinfo {editor}
  {edited by\ \bibinfo {editor} {\bibfnamefont {R.}~\bibnamefont {Broglia}}\
  and\ \bibinfo {editor} {\bibfnamefont {J.~R.}\ \bibnamefont {Schrieffer}}}\
  (\bibinfo  {publisher} {North Holland},\ \bibinfo {address} {Amsterdam},\
  \bibinfo {year} {1994})\ pp.\ \bibinfo {pages} {5--30}\BibitemShut {NoStop}%
\bibitem [{\citenamefont {Schulz}(1990)}]{Schultz90}%
  \BibitemOpen
  \bibfield  {author} {\bibinfo {author} {\bibfnamefont {H.~J.}\ \bibnamefont
  {Schulz}},\ }\href@noop {} {\bibfield  {journal} {\bibinfo  {journal} {Phys.
  Rev. Lett.}\ }\textbf {\bibinfo {volume} {64}},\ \bibinfo {pages} {2831}
  (\bibinfo {year} {1990})}\BibitemShut {NoStop}%
\bibitem [{\citenamefont {Frahm}\ and\ \citenamefont
  {Korepin}(1990)}]{Frahm90}%
  \BibitemOpen
  \bibfield  {author} {\bibinfo {author} {\bibfnamefont {H.}~\bibnamefont
  {Frahm}}\ and\ \bibinfo {author} {\bibfnamefont {V.~E.}\ \bibnamefont
  {Korepin}},\ }\href@noop {} {\bibfield  {journal} {\bibinfo  {journal} {Phys.
  Rev. B}\ }\textbf {\bibinfo {volume} {42}},\ \bibinfo {pages} {10553}
  (\bibinfo {year} {1990})}\BibitemShut {NoStop}%
\bibitem [{\citenamefont {Kawakami}\ and\ \citenamefont
  {Yang}(1990)}]{Kawakami90}%
  \BibitemOpen
  \bibfield  {author} {\bibinfo {author} {\bibfnamefont {N.}~\bibnamefont
  {Kawakami}}\ and\ \bibinfo {author} {\bibfnamefont {S.-K.}\ \bibnamefont
  {Yang}},\ }\href@noop {} {\bibfield  {journal} {\bibinfo  {journal} {Phys.
  Lett. A}\ }\textbf {\bibinfo {volume} {148}},\ \bibinfo {pages} {359}
  (\bibinfo {year} {1990})}\BibitemShut {NoStop}%
\bibitem [{\citenamefont {Penc}\ and\ \citenamefont
  {S{\'o}lyom}(1991)}]{Penc91}%
  \BibitemOpen
  \bibfield  {author} {\bibinfo {author} {\bibfnamefont {K.}~\bibnamefont
  {Penc}}\ and\ \bibinfo {author} {\bibfnamefont {J.}~\bibnamefont
  {S{\'o}lyom}},\ }\href@noop {} {\bibfield  {journal} {\bibinfo  {journal}
  {Phys. Rev. B}\ }\textbf {\bibinfo {volume} {44}},\ \bibinfo {pages} {12690}
  (\bibinfo {year} {1991})}\BibitemShut {NoStop}%
\bibitem [{\citenamefont {Bethe}(1931)}]{Bethe31}%
  \BibitemOpen
  \bibfield  {author} {\bibinfo {author} {\bibfnamefont {H.}~\bibnamefont
  {Bethe}},\ }\href@noop {} {\bibfield  {journal} {\bibinfo  {journal} {Z.
  Physik}\ }\textbf {\bibinfo {volume} {71}},\ \bibinfo {pages} {205} (\bibinfo
  {year} {1931})}\BibitemShut {NoStop}%
\bibitem [{\citenamefont {Gaudin}(1967)}]{Gaudin67}%
  \BibitemOpen
  \bibfield  {author} {\bibinfo {author} {\bibfnamefont {M.}~\bibnamefont
  {Gaudin}},\ }\href@noop {} {\bibfield  {journal} {\bibinfo  {journal} {Phys.
  Lett. A}\ }\textbf {\bibinfo {volume} {24}},\ \bibinfo {pages} {55} (\bibinfo
  {year} {1967})}\BibitemShut {NoStop}%
\bibitem [{\citenamefont {Yang}(1967)}]{Yang67}%
  \BibitemOpen
  \bibfield  {author} {\bibinfo {author} {\bibfnamefont {C.~N.}\ \bibnamefont
  {Yang}},\ }\href@noop {} {\bibfield  {journal} {\bibinfo  {journal} {Phys.
  Rev. Lett.}\ }\textbf {\bibinfo {volume} {19}},\ \bibinfo {pages} {1312}
  (\bibinfo {year} {1967})}\BibitemShut {NoStop}%
\bibitem [{\citenamefont {Gaudin}(2014)}]{Gaudin_book}%
  \BibitemOpen
  \bibfield  {author} {\bibinfo {author} {\bibfnamefont {M.}~\bibnamefont
  {Gaudin}},\ }\href@noop {} {\emph {\bibinfo {title} {The Bethe
  Wavefunction}}}\ (\bibinfo  {publisher} {Cambridge University Press},\
  \bibinfo {address} {Cambridge},\ \bibinfo {year} {2014})\BibitemShut
  {NoStop}%
\bibitem [{\citenamefont {Essler}\ \emph {et~al.}(2005)\citenamefont {Essler},
  \citenamefont {Frahm}, \citenamefont {G{\"o}hmann}, \citenamefont
  {Kl{\"u}mper},\ and\ \citenamefont {Korepin}}]{Essler_book}%
  \BibitemOpen
  \bibfield  {author} {\bibinfo {author} {\bibfnamefont {F.~H.~L.}\
  \bibnamefont {Essler}}, \bibinfo {author} {\bibfnamefont {H.}~\bibnamefont
  {Frahm}}, \bibinfo {author} {\bibfnamefont {F.}~\bibnamefont {G{\"o}hmann}},
  \bibinfo {author} {\bibfnamefont {A.}~\bibnamefont {Kl{\"u}mper}},\ and\
  \bibinfo {author} {\bibfnamefont {V.~E.}\ \bibnamefont {Korepin}},\
  }\href@noop {} {\emph {\bibinfo {title} {The One-Dimensional Hubbard
  Model}}}\ (\bibinfo  {publisher} {Cambridge University Press},\ \bibinfo
  {address} {Cambridge},\ \bibinfo {year} {2005})\BibitemShut {NoStop}%
\bibitem [{\citenamefont {Slavnov}(1989)}]{Slavnov89}%
  \BibitemOpen
  \bibfield  {author} {\bibinfo {author} {\bibfnamefont {N.~A.}\ \bibnamefont
  {Slavnov}},\ }\href@noop {} {\bibfield  {journal} {\bibinfo  {journal}
  {Theor. Math. Phys.}\ }\textbf {\bibinfo {volume} {79}},\ \bibinfo {pages}
  {502} (\bibinfo {year} {1989})}\BibitemShut {NoStop}%
\bibitem [{\citenamefont {Korepin}(1982)}]{Korepin82}%
  \BibitemOpen
  \bibfield  {author} {\bibinfo {author} {\bibfnamefont {V.~E.}\ \bibnamefont
  {Korepin}},\ }\href@noop {} {\bibfield  {journal} {\bibinfo  {journal}
  {Commun. Math. Phys.}\ }\textbf {\bibinfo {volume} {86}},\ \bibinfo {pages}
  {391} (\bibinfo {year} {1982})}\BibitemShut {NoStop}%
\bibitem [{\citenamefont {Drinfeld}(1983)}]{Drinfeld83}%
  \BibitemOpen
  \bibfield  {author} {\bibinfo {author} {\bibfnamefont {V.~G.}\ \bibnamefont
  {Drinfeld}},\ }\href@noop {} {\bibfield  {journal} {\bibinfo  {journal} {Sov.
  Math. Dokl.}\ }\textbf {\bibinfo {volume} {28}},\ \bibinfo {pages} {667}
  (\bibinfo {year} {1983})}\BibitemShut {NoStop}%
\end{thebibliography}%


\begin{thebibliography}{16}%
\makeatletter
\providecommand \@ifxundefined [1]{%
 \@ifx{#1\undefined}
}%
\providecommand \@ifnum [1]{%
 \ifnum #1\expandafter \@firstoftwo
 \else \expandafter \@secondoftwo
 \fi
}%
\providecommand \@ifx [1]{%
 \ifx #1\expandafter \@firstoftwo
 \else \expandafter \@secondoftwo
 \fi
}%
\providecommand \natexlab [1]{#1}%
\providecommand \enquote  [1]{``#1''}%
\providecommand \bibnamefont  [1]{#1}%
\providecommand \bibfnamefont [1]{#1}%
\providecommand \citenamefont [1]{#1}%
\providecommand \href@noop [0]{\@secondoftwo}%
\providecommand \href [0]{\begingroup \@sanitize@url \@href}%
\providecommand \@href[1]{\@@startlink{#1}\@@href}%
\providecommand \@@href[1]{\endgroup#1\@@endlink}%
\providecommand \@sanitize@url [0]{\catcode `\\12\catcode `\$12\catcode
  `\&12\catcode `\#12\catcode `\^12\catcode `\_12\catcode `\%12\relax}%
\providecommand \@@startlink[1]{}%
\providecommand \@@endlink[0]{}%
\providecommand \url  [0]{\begingroup\@sanitize@url \@url }%
\providecommand \@url [1]{\endgroup\@href {#1}{\urlprefix }}%
\providecommand \urlprefix  [0]{URL }%
\providecommand \Eprint [0]{\href }%
\providecommand \doibase [0]{https://doi.org/}%
\providecommand \selectlanguage [0]{\@gobble}%
\providecommand \bibinfo  [0]{\@secondoftwo}%
\providecommand \bibfield  [0]{\@secondoftwo}%
\providecommand \translation [1]{[#1]}%
\providecommand \BibitemOpen [0]{}%
\providecommand \bibitemStop [0]{}%
\providecommand \bibitemNoStop [0]{.\EOS\space}%
\providecommand \EOS [0]{\spacefactor3000\relax}%
\providecommand \BibitemShut  [1]{\csname bibitem#1\endcsname}%
\let\auto@bib@innerbib\@empty
\bibitem [{\citenamefont {Lieb}\ and\ \citenamefont {Wu}(1968)}]{LiebWu68}%
  \BibitemOpen
  \bibfield  {author} {\bibinfo {author} {\bibfnamefont {E.~H.}\ \bibnamefont
  {Lieb}}\ and\ \bibinfo {author} {\bibfnamefont {F.~Y.}\ \bibnamefont {Wu}},\
  }\href@noop {} {\bibfield  {journal} {\bibinfo  {journal} {Phys. Rev. Lett.}\
  }\textbf {\bibinfo {volume} {20}},\ \bibinfo {pages} {1445} (\bibinfo {year}
  {1968})}\BibitemShut {NoStop}%
\bibitem [{\citenamefont {Bethe}(1931)}]{Bethe31}%
  \BibitemOpen
  \bibfield  {author} {\bibinfo {author} {\bibfnamefont {H.}~\bibnamefont
  {Bethe}},\ }\href@noop {} {\bibfield  {journal} {\bibinfo  {journal} {Z.
  Physik}\ }\textbf {\bibinfo {volume} {71}},\ \bibinfo {pages} {205} (\bibinfo
  {year} {1931})}\BibitemShut {NoStop}%
\bibitem [{\citenamefont {Gaudin}(1967)}]{Gaudin67}%
  \BibitemOpen
  \bibfield  {author} {\bibinfo {author} {\bibfnamefont {M.}~\bibnamefont
  {Gaudin}},\ }\href@noop {} {\bibfield  {journal} {\bibinfo  {journal} {Phys.
  Lett. A}\ }\textbf {\bibinfo {volume} {24}},\ \bibinfo {pages} {55} (\bibinfo
  {year} {1967})}\BibitemShut {NoStop}%
\bibitem [{\citenamefont {Yang}(1967)}]{Yang67}%
  \BibitemOpen
  \bibfield  {author} {\bibinfo {author} {\bibfnamefont {C.~N.}\ \bibnamefont
  {Yang}},\ }\href@noop {} {\bibfield  {journal} {\bibinfo  {journal} {Phys.
  Rev. Lett.}\ }\textbf {\bibinfo {volume} {19}},\ \bibinfo {pages} {1312}
  (\bibinfo {year} {1967})}\BibitemShut {NoStop}%
\bibitem [{\citenamefont {Ogata}\ and\ \citenamefont {Shiba}(1990)}]{Ogata90}%
  \BibitemOpen
  \bibfield  {author} {\bibinfo {author} {\bibfnamefont {M.}~\bibnamefont
  {Ogata}}\ and\ \bibinfo {author} {\bibfnamefont {H.}~\bibnamefont {Shiba}},\
  }\href@noop {} {\bibfield  {journal} {\bibinfo  {journal} {Phys. Rev. B}\
  }\textbf {\bibinfo {volume} {41}},\ \bibinfo {pages} {2326} (\bibinfo {year}
  {1990})}\BibitemShut {NoStop}%
\bibitem [{\citenamefont {Gaudin}(2014)}]{Gaudin_book}%
  \BibitemOpen
  \bibfield  {author} {\bibinfo {author} {\bibfnamefont {M.}~\bibnamefont
  {Gaudin}},\ }\href@noop {} {\emph {\bibinfo {title} {The Bethe
  Wavefunction}}}\ (\bibinfo  {publisher} {Cambridge University Press},\
  \bibinfo {address} {Cambridge},\ \bibinfo {year} {2014})\BibitemShut
  {NoStop}%
\bibitem [{\citenamefont {Essler}\ \emph {et~al.}(2005)\citenamefont {Essler},
  \citenamefont {Frahm}, \citenamefont {G{\"o}hmann}, \citenamefont
  {Kl{\"u}mper},\ and\ \citenamefont {Korepin}}]{Essler_book}%
  \BibitemOpen
  \bibfield  {author} {\bibinfo {author} {\bibfnamefont {F.~H.~L.}\
  \bibnamefont {Essler}}, \bibinfo {author} {\bibfnamefont {H.}~\bibnamefont
  {Frahm}}, \bibinfo {author} {\bibfnamefont {F.}~\bibnamefont {G{\"o}hmann}},
  \bibinfo {author} {\bibfnamefont {A.}~\bibnamefont {Kl{\"u}mper}},\ and\
  \bibinfo {author} {\bibfnamefont {V.~E.}\ \bibnamefont {Korepin}},\
  }\href@noop {} {\emph {\bibinfo {title} {The One-Dimensional Hubbard
  Model}}}\ (\bibinfo  {publisher} {Cambridge University Press},\ \bibinfo
  {address} {Cambridge},\ \bibinfo {year} {2005})\BibitemShut {NoStop}%
\bibitem [{\citenamefont {Sklyanin}\ \emph {et~al.}(1979)\citenamefont
  {Sklyanin}, \citenamefont {Takhtajan},\ and\ \citenamefont
  {Faddeev}}]{Faddeev79}%
  \BibitemOpen
  \bibfield  {author} {\bibinfo {author} {\bibfnamefont {E.~K.}\ \bibnamefont
  {Sklyanin}}, \bibinfo {author} {\bibfnamefont {L.~A.}\ \bibnamefont
  {Takhtajan}},\ and\ \bibinfo {author} {\bibfnamefont {L.~D.}\ \bibnamefont
  {Faddeev}},\ }\href@noop {} {\bibfield  {journal} {\bibinfo  {journal}
  {Theor. Math. Phys.}\ }\textbf {\bibinfo {volume} {40}},\ \bibinfo {pages}
  {688} (\bibinfo {year} {1979})}\BibitemShut {NoStop}%
\bibitem [{\citenamefont {Korepin}\ \emph {et~al.}(1993)\citenamefont
  {Korepin}, \citenamefont {Bogoliubov},\ and\ \citenamefont
  {Izergin}}]{Korepin_book}%
  \BibitemOpen
  \bibfield  {author} {\bibinfo {author} {\bibfnamefont {V.~E.}\ \bibnamefont
  {Korepin}}, \bibinfo {author} {\bibfnamefont {N.~M.}\ \bibnamefont
  {Bogoliubov}},\ and\ \bibinfo {author} {\bibfnamefont {A.~G.}\ \bibnamefont
  {Izergin}},\ }\href@noop {} {\emph {\bibinfo {title} {Quantum inverse
  scattering methods and correlation functions}}}\ (\bibinfo  {publisher}
  {Cambridge University Press},\ \bibinfo {year} {1993})\BibitemShut {NoStop}%
\bibitem [{\citenamefont {Slavnov}(1989)}]{Slavnov89}%
  \BibitemOpen
  \bibfield  {author} {\bibinfo {author} {\bibfnamefont {N.~A.}\ \bibnamefont
  {Slavnov}},\ }\href@noop {} {\bibfield  {journal} {\bibinfo  {journal}
  {Theor. Math. Phys.}\ }\textbf {\bibinfo {volume} {79}},\ \bibinfo {pages}
  {502} (\bibinfo {year} {1989})}\BibitemShut {NoStop}%
\bibitem [{\citenamefont {Gaudin}\ \emph {et~al.}(1981)\citenamefont {Gaudin},
  \citenamefont {McCoy},\ and\ \citenamefont {Wu}}]{Gaudin81}%
  \BibitemOpen
  \bibfield  {author} {\bibinfo {author} {\bibfnamefont {M.}~\bibnamefont
  {Gaudin}}, \bibinfo {author} {\bibfnamefont {B.~M.}\ \bibnamefont {McCoy}},\
  and\ \bibinfo {author} {\bibfnamefont {T.~T.}\ \bibnamefont {Wu}},\
  }\href@noop {} {\bibfield  {journal} {\bibinfo  {journal} {Phys. Rev. D}\
  }\textbf {\bibinfo {volume} {23}},\ \bibinfo {pages} {417} (\bibinfo {year}
  {1981})}\BibitemShut {NoStop}%
\bibitem [{\citenamefont {Korepin}(1982)}]{Korepin82}%
  \BibitemOpen
  \bibfield  {author} {\bibinfo {author} {\bibfnamefont {V.~E.}\ \bibnamefont
  {Korepin}},\ }\href@noop {} {\bibfield  {journal} {\bibinfo  {journal}
  {Commun. Math. Phys.}\ }\textbf {\bibinfo {volume} {86}},\ \bibinfo {pages}
  {391} (\bibinfo {year} {1982})}\BibitemShut {NoStop}%
\bibitem [{\citenamefont {Penc}\ \emph {et~al.}(1996)\citenamefont {Penc},
  \citenamefont {Hallberg}, \citenamefont {Mila},\ and\ \citenamefont
  {Shiba}}]{Penc96}%
  \BibitemOpen
  \bibfield  {author} {\bibinfo {author} {\bibfnamefont {K.}~\bibnamefont
  {Penc}}, \bibinfo {author} {\bibfnamefont {K.}~\bibnamefont {Hallberg}},
  \bibinfo {author} {\bibfnamefont {F.}~\bibnamefont {Mila}},\ and\ \bibinfo
  {author} {\bibfnamefont {H.}~\bibnamefont {Shiba}},\ }\href@noop {}
  {\bibfield  {journal} {\bibinfo  {journal} {Phys. Rev. Lett.}\ }\textbf
  {\bibinfo {volume} {77}},\ \bibinfo {pages} {1390} (\bibinfo {year}
  {1996})}\BibitemShut {NoStop}%
\bibitem [{\citenamefont {Drinfeld}(1983)}]{Drinfeld83}%
  \BibitemOpen
  \bibfield  {author} {\bibinfo {author} {\bibfnamefont {V.~G.}\ \bibnamefont
  {Drinfeld}},\ }\href@noop {} {\bibfield  {journal} {\bibinfo  {journal} {Sov.
  Math. Dokl.}\ }\textbf {\bibinfo {volume} {28}},\ \bibinfo {pages} {667}
  (\bibinfo {year} {1983})}\BibitemShut {NoStop}%
\bibitem [{\citenamefont {Kitanine}\ \emph {et~al.}(1999)\citenamefont
  {Kitanine}, \citenamefont {Maillet},\ and\ \citenamefont
  {Tetras}}]{Kitanine99}%
  \BibitemOpen
  \bibfield  {author} {\bibinfo {author} {\bibfnamefont {N.}~\bibnamefont
  {Kitanine}}, \bibinfo {author} {\bibfnamefont {J.}~\bibnamefont {Maillet}},\
  and\ \bibinfo {author} {\bibfnamefont {V.}~\bibnamefont {Tetras}},\
  }\href@noop {} {\bibfield  {journal} {\bibinfo  {journal} {Nucl. Phys. B}\
  }\textbf {\bibinfo {volume} {554}},\ \bibinfo {pages} {647} (\bibinfo {year}
  {1999})}\BibitemShut {NoStop}%
\bibitem [{\citenamefont {Kitanine}\ and\ \citenamefont
  {Maillet}(2000)}]{Kitanine00}%
  \BibitemOpen
  \bibfield  {author} {\bibinfo {author} {\bibfnamefont {N.}~\bibnamefont
  {Kitanine}}\ and\ \bibinfo {author} {\bibfnamefont {J.~M.}\ \bibnamefont
  {Maillet}},\ }\href@noop {} {\bibfield  {journal} {\bibinfo  {journal} {Nucl.
  Phys. B}\ }\textbf {\bibinfo {volume} {567}} (\bibinfo {year}
  {2000})}\BibitemShut {NoStop}%
\end{thebibliography}%

\end{document}


\title{\noindent Supplementary material for\\ Splitting of Fermi point of strongly interacting electrons in one dimension:\\ A nonlinear effect of spin-charge separation}
\author{O. Tsyplyatyev}
\maketitle

\section{Lieb-Wu solution}

The many-particle eigenstates of 1D Fermi-Hubbard model in Eq.~(1)
of the main text were constructed in \citep{LiebWu68}. In the representation
of second quantisation, $\left|\Psi\right\rangle =\sum_{\mathbf{j},\boldsymbol{\alpha}}\psi_{\mathbf{j}\boldsymbol{\alpha}}c_{j_{1}\alpha_{1}}^{\dagger}\cdots c_{j_{N}\alpha_{N}}^{\dagger}\left|0\right\rangle $,
they are described by the amplitude $\psi_{\mathbf{j}\boldsymbol{\alpha}}$
of finding all $N$ particles at a given set of sites $j_{1}\dots j_{N}=\mathbf{j}$
and with a given configuration of their spins $\alpha_{1}\dots\alpha_{M}=\boldsymbol{\alpha}$.
These amplitudes have the form of superposition of plane waves according
to the Bethe's hypothesis \citep{Bethe31},
\begin{equation}
\psi_{\mathbf{j}\boldsymbol{\alpha}}=\sum_{Q}A_{QO\alpha}e^{iQ\mathbf{k}\cdot O\mathbf{j}},\label{eq:Psi_LW}
\end{equation}
where $O$ is the permutation that orders all $N$ coordinates such
that 
\begin{equation}
Oj_{1}<\dots<Oj_{N},
\end{equation}
the (charge) momenta of $N$ particles are $\mathbf{k}=k_{1},\dots k_{N}$,
and $\sum_{Q}$ is the sum over all permutations of $N$ momenta $k_{j}$.
The phase $A_{PQ\alpha}$ in this superposition is neither $1$ nor
$-1$, and it depends on the configuration of all $M$ spins $O\alpha$
explicitly. The algebraic method of constructing it was proposed in
\citep{Gaudin67,Yang67} in the form of Bethe ansatz, producing another
``nested'' Bethe-ansatz wave function for the 1D Fermi Hubbard model,

\begin{equation}
A_{QO\boldsymbol{\alpha}}=\left(-1\right)^{QO}\sum_{R}\left(\prod_{1\leq l<m<M}\frac{R\lambda_{l}-R\lambda_{m}-\frac{iU}{2t}}{R\lambda_{l}-R\lambda_{m}}\right)\prod_{l}^{M}\frac{\frac{iU}{2t}}{R\lambda_{l}-\sin Q\lambda_{l}+\frac{iU}{4t}}\sum_{j=1}^{x_{l}-1}\frac{R\lambda_{l}-\sin Q\lambda_{j}-\frac{iU}{4t}}{R\lambda_{l}-\sin Q\lambda_{j}-\frac{iU}{4t}}\label{eq:AQOalpha}
\end{equation}
where $x_{1},\dots,x_{M}=\mathbf{x}$ are the coordinates of $M$
spins $\uparrow$ in the configuration $O\boldsymbol{\alpha}$ of
all spins of $N$ particles, $\lambda_{1},\dots,\lambda_{M}=\boldsymbol{\lambda}$
are the spin momenta associated with these $M$ spins $\uparrow$,
and $\sum_{R}$ is the sum over all permutations of these spin momenta.

The momenta are quantised by boundary conditions. Application of the
periodic boundary condition to the many-particle wave function in
Eq.~(\ref{eq:Psi_LW}) gives the Lieb-Wu equations \citep{LiebWu68},
\begin{align}
k_{j}L-\sum_{m=1}^{M}\varphi\left(\lambda_{m}-\sin k_{j}\right) & =2\pi I_{j},\label{eq:LW_charge_arbU}\\
\sum_{j=1}^{N}\varphi\left(\lambda_{m}-\sin k_{j}\right)-\sum_{l=1}^{M}\varphi\left(\lambda_{m}/2-\lambda_{l}/2\right) & =2\pi J_{m},\label{eq:LW_spin_arbU}\\
\mathrm{with}\quad\varphi\left(x\right)=-2\arctan\left(\frac{4tx}{U}\right),
\end{align}
where $N$ non-equal integers $I_{j}$ and $M$ non-equal integers
$J_{m}$ define the solution for the charge $k_{j}$ and the spin
$\lambda_{j}$ momenta for a given value of the interaction strength
$U/t$. This solution gives the eigenenergy of the many-particle state
as $E=t\sum_{j=1}^{N}\cos k_{j}$ and its momentum as $P=\sum_{j=1}^{N}k_{j}$.
These simultaneous quantisation conditions for all spin and charge
degrees of freedom are a system of $N+M$ connected equations for
any finite $U$.

\subsection{$U=\infty$ limit}

In the regime of strong coupling $U/t\gg1$, it was noted in \citep{Ogata90}
that the spin momenta $q_{m}$ in the solutions of Eqs.~(\ref{eq:LW_charge_arbU},\ref{eq:LW_spin_arbU})
become large (growing with $U/t$ without constraint for large $U$)
while the charge momenta $k_{j}$ remain finite in the limit. This
property allows to decouple the system of the $N+M$ connected equations
into, at least, some disconnected parts by means of $t/U\ll1$ expansion,
simplifying the solution in this $U=\infty$ limit. It is convenient
to perform such an expansion by introducing the following mapping
for the spin degrees of freedom,

\begin{equation}
\lambda_{l}=-\frac{iU}{4t}\frac{e^{iq_{l}}+1}{e^{iq_{l}}-1},\label{eq:q_mapping_Uinf}
\end{equation}
where, in addition to the rescaling by $U/t$, the parameterisation
of the spin momenta is changed from Orbach \citep{Gaudin_book} to
coordinate so the representation for the charge and spin momenta is
the same. 

Under substitution of the mapping in Eq.~(\ref{eq:q_mapping_Uinf})
the leading $t/U$ order term in the Taylor expansion of Eqs.~(\ref{eq:LiebWu_charge_SM},\ref{eq:LiebWu_spin_SM})
becomes 

\begin{align}
Lk_{j} & -\sum_{m}q_{m}=2\pi I_{j},\label{eq:LiebWu_charge_SM}\\
Nq_{m} & -2\sum_{l\neq m}^{M}\varphi_{lm}=2\pi J_{m},\label{eq:LiebWu_spin_SM}
\end{align}
where 
\begin{equation}
e^{i2\varphi_{lm}}=-\frac{e^{iq_{l}+iq_{m}}+1-2e^{iq_{l}}}{e^{iq_{l}+iq_{m}}+1-2e^{iq_{m}}}\label{eq:2body_scattering_phase_AFM}
\end{equation}
are the two-spinon scattering phases. Note that the mapping in Eq.~(\ref{eq:q_mapping_Uinf})
is chosen such so the resulting two-spinon scattering phases above
correspond to the antiferromagnetic case since the Hubbard model in
the strong coupling regime is well-approximated by the $t-J$ model
\citep{Essler_book}, with its antiferromagnetic correlations for
the spin degrees of freedom making this choice the most natural. Taking
additionally the low density limit, $N/L\ll1$, leaves the Lieb-Wu
equations in Eqs.~(\ref{eq:LiebWu_charge_SM},\ref{eq:LiebWu_spin_SM})
unchanged but simplifies the eigenenergy of the many-particle state
as $E=t\sum_{j=1}^{N}k_{j}^{2}/2$. 

As a result of taking this limit, the spin part of the Lieb-Wu equations
decouples completely in Eq.~(\ref{eq:LiebWu_spin_SM}) becoming
a self-contained set of $M$ nonlinear equations, which are exactly
the Bethe equations for the antiferromagnetic Heisenberg chain \citep{Bethe31}.
Once this system of $M$ equations for $q_{m}$ is solved, each equation
for $k_{j}$ in Eq.~(\ref{eq:LiebWu_charge_SM}) becomes just an
independent single-particle quantisation condition.

The form of the Lieb-Wu eigenstates in Eq.~(\ref{eq:Psi_LW}) also
simplifies in the $U=\infty$ limit. Under the substitution of the
mapping in Eq.~(\ref{eq:q_mapping_Uinf}) into the phase factor
$A_{QO\boldsymbol{\alpha}}$ in Eq.~(\ref{eq:AQOalpha}), the leading
order term of the Taylor expansion in $t/U$ of the resulting expression
is 

\begin{equation}
A_{QO\boldsymbol{\alpha}}=\left(-1\right)^{QO}\sum_{R}e^{i\sum_{l<m}^{M}\varphi_{R_{l},R_{m}}+\sum_{l=1}^{M}q_{R_{l}}x_{l}}.\label{eq:AQOalpha_Uinf}
\end{equation}
Here a factor that depends only on the spin momenta $q_{j}$ but not
on any permutation $Q$ or $R$ was ignored since it can be absorbed
by the normalisation of the whole wave function $\left|\Psi\right\rangle $.

The resulting expression under the sum over $R$ in Eq.~(\ref{eq:AQOalpha_Uinf})
depends only on the spin momenta $q_{j}$ and spin coordinates $x_{l}$,
which allows to factorise the wave function in Eq.~(\ref{eq:Psi_LW})
as \citep{Ogata90} 

\begin{equation}
\psi_{\mathbf{j}\boldsymbol{\alpha}}=\psi_{\mathbf{j}}^{c}\cdot\psi_{\boldsymbol{\alpha}}^{s},
\end{equation}
where the spin part is 
\begin{equation}
\psi_{\boldsymbol{\alpha}}^{s}=\sum_{R}e^{i\sum_{l<m}\varphi_{R_{l},R_{m}}+\sum_{l=1}^{M}q_{R_{l}}x_{l}}\label{eq:Psi_s_SM}
\end{equation}
and the charge part, 
\begin{equation}
\psi_{\mathbf{j}}^{c}=\sum_{Q}\left(-1\right)^{Q}e^{iQ\mathbf{k}\cdot\mathbf{j}},\label{eq:Psi_c_SM}
\end{equation}
depends only on the charge momenta $k_{j}$ and charge coordinates
$j_{l}$. The ordering permutation $O$ disappears from this limiting
expression since the sum over $Q$ in Eq.~(\ref{eq:Psi_c_SM}) is
a Slater determinant, under which exchange of the rows corresponding
to the charge coordinates $x_{j}$ cancels the $\left(-1\right)^{O}$
factor altogether. 

The expressions in Eqs.~(\ref{eq:LiebWu_charge_SM},\ref{eq:LiebWu_spin_SM})
and in Eqs.~(\ref{eq:Psi_s_SM},\ref{eq:Psi_c_SM}) are presented
in Eqs.~(2-5) of the main text as the starting point for the calculation
in this work.

\section{Algebraic representation of Bethe ansatz}

The Bethe states for the spin part of the wave function in Eq.~(\ref{eq:Psi_s_SM})
are the eigenstates of the antiferromangetic Heisenberg model \citep{Bethe31}.
In Eq.~(\ref{eq:Psi_s_SM}) these many-body states are written in
the so-called coordinate representation, in which they are not factorisable,
making calculations of scalar products and expectation values almost
intractable. Another, the so-called algebraic representation of Bethe
ansatz was invented \citep{Faddeev79} as a way for solving this kind
of problems. In the algebraic representation the Bethe states are
factorised in terms of operators with given commutation rations, which
can be used to perform practical calculations.

It is more convenient to construct such an algebraic representation
for a bit more general spin model, the XXZ model
\begin{equation}
H=\sum_{j=1}^{N}\left(\frac{S_{j}^{+}S_{j+1}^{-}+S_{j}^{+}S_{j+1}^{-}}{2}+\Delta S_{j}^{z}S_{j+1}^{z}\right).\label{eq:H_XXZ}
\end{equation}
The eigenstates of this model are the same as in Eq.~(\ref{eq:Psi_s_SM})
where the $M$ spin momenta $q_{j}$ satisfying the spin part of Lieb-Wu
equation (\ref{eq:LiebWu_spin_SM}) with the two-spinon scattering
phase 
\begin{equation}
e^{i2\varphi_{lm}}=-\frac{e^{iq_{l}+iq_{m}}+1-2\Delta e^{iq_{l}}}{e^{iq_{l}+iq_{m}}+1-2\Delta e^{iq_{m}}}.\label{eq:2body_scattering_phase_XXZ}
\end{equation}
For $\Delta=1$ Eqs.~(\ref{eq:H_XXZ},\ref{eq:2body_scattering_phase_XXZ})
are the antiferromagnetic Heisenberg model, $H=\sum_{i}\mathbf{S}_{i}\cdot\mathbf{S}_{i+1}$,
and its two-spinon scattering phase in Eq.~(\ref{eq:2body_scattering_phase_AFM}).

Following the notations of the book in \citep{Korepin_book}, the
many-body wave function of the XXZ model can be written using (Bethe
ansatz) operators that satisfy an algebra generated by the Yang-Baxter
equation as 

\begin{equation}
\left|\mathbf{u}\right\rangle =\prod_{j=1}^{M}C\left(u_{j}\right)\left|\Downarrow\right\rangle ,\label{eq:ABA_state}
\end{equation}
where $u_{j}$ are $M$ complex parameters corresponding to $M$ spin
momenta $q_{j}$, $\left|\Downarrow\right\rangle $ is the ``vacuum''
state (associated with all the $N$ spins on the chain are in the
down state configuration), and $C\left(u\right)$ is one of the four
matrix elements of the transition (monodromy) matrix 

\begin{equation}
T\left(u\right)=\left(\begin{array}{cc}
A\left(u\right) & B\left(u\right)\\
C\left(u\right) & D\left(u\right)
\end{array}\right).\label{eq:Tmatrix_def}
\end{equation}
This matrix is defined in an auxiliary $2\times2$ space and is a
function of the parameter $u$ that can be arbitrary complex number.
This $T$-matrix is a solution of the Yang-Baxter equation
\begin{equation}
R\left(u-v\right)\left(T\left(u\right)\otimes T\left(v\right)\right)=\left(T\left(v\right)\otimes T\left(u\right)\right)R\left(u-v\right),\label{eq:YangBaxter_eq}
\end{equation}
ensuring that any more than two-body scattering matrix factorises
into a product of only two-body scattering matrices. 

The so-called $R$-matrix here acts on a $4\times4$ tensor product
$V_{1}\otimes V_{2}$ space, where $V_{1}$ and $V_{2}$ are two-element
subspaces, each of which consists of two spin states $\left|\downarrow\right\rangle _{j}$
and $\left|\uparrow\right\rangle _{j}$. For the model in Eq.~(\ref{eq:H_XXZ})
the $R$-matrix is \citep{Korepin_book}
\begin{equation}
R\left(u\right)=\left(\begin{array}{cccc}
1\\
 & b\left(u\right) & c\left(u\right)\\
 & c\left(u\right) & b\left(u\right)\\
 &  &  & 1
\end{array}\right),\label{eq:Rmatrix_XXZ}
\end{equation}
where 
\begin{equation}
b\left(u\right)=\frac{\sinh\left(u\right)}{\sinh\left(u+2\eta\right)},\quad c\left(u\right)=\frac{\sinh\left(2\eta\right)}{\sinh\left(u+2\eta\right)},\label{eq:bc_XXZ_def}
\end{equation}
and $\eta$ is a real number corresponding to the interaction strength.
Note that this $R$-matrix also satisfies the Yang-Baxter equation,
$R_{12}\left(u_{1}-u_{2}\right)R_{13}\left(u_{1}\right)R_{23}\left(u_{2}\right)=R_{23}\left(u_{2}\right)R_{13}\left(u_{1}\right)R_{12}\left(u_{1}-u_{2}\right)$.

A two-element subspace of the $R$-matrix can be identified with the
two-state spin subspace on the lattice site $j$ in the model in Eq.~(\ref{eq:H_XXZ}).
Then, the quantum version of the so-called Lax matrix ($L$-matrix)
for a single spin site $j$ can be defined as $L_{j}=R_{1j}$ \citep{Faddeev79},
where the subspace $1$ plays the role of the auxiliary $2\times2$
space, in which the $T$-matrix in Eq.~(\ref{eq:Tmatrix_def}) is
defined. In this auxiliary space the matrix form of the Lax operator
is
\begin{equation}
L_{j}\left(u\right)=\left(\begin{array}{cc}
\frac{\cosh\left(u+\eta2S_{j}^{z}\right)}{\cosh\left(u-\eta\right)} & -i\frac{\sinh\left(2\eta\right)S_{j}^{-}}{\cosh\left(u-\eta\right)}\\
-i\frac{\sinh\left(2\eta\right)S_{j}^{+}}{\cosh\left(u-\eta\right)} & \frac{\cosh\left(u-\eta2S_{j}^{z}\right)}{\cosh\left(u-\eta\right)}
\end{array}\right).\label{eq:Lmatrix_XXZ}
\end{equation}
The prefactor in front of $L_{j}\left(u\right)$, and the matrix elements
of the $R$-matrix in Eq.~(\ref{eq:Rmatrix_XXZ}), are chosen such
that in the non-interacting limit $\eta=0$ the $L$-operator is a
unit matrix. Lastly, the transition matrix $T\left(u\right)$ for
a chain consisting of $N$ spin sites can be constructed similarly
to the Lax method for classical systems as 

\begin{equation}
T\left(u\right)=\sum_{j=1}^{N}L_{j}\left(u\right),\label{eq:Tmatrix_Lax_construction}
\end{equation}
providing a definition of the algebraic Bethe ansatz operators in
terms of the spin operators of the model in Eq.~(\ref{eq:H_XXZ}).
Note that starting from the $L$-matrix in Eq.~(\ref{eq:Lmatrix_XXZ}),
which satisfy the Yang-Baxter equation by construction, it can be
shown explicitly that the $T$-matrix defined in Eq.~(\ref{eq:Tmatrix_Lax_construction})
also satisfies the same Yang-Baxter equation, \emph{e.g. }see proof
in \citep{Korepin_book}. 

The $16$ entries (in the $4\times4$ space) of Yang-Baxter equation
(\ref{eq:YangBaxter_eq}) with the $R$-matrix in Eq.~(\ref{eq:Rmatrix_XXZ})
give the commutation relations between all four Bethe ansatz operators
$A\left(u\right),B\left(u\right),C\left(u\right),$ and $D\left(u\right)$,
which are the the matrix elements of $T$-matrix. The explicit from
for the four of them that will be needed later is 
\begin{equation}
\left[B_{u},C_{v}\right]=\frac{c\left(u-v\right)}{b\left(u-v\right)}\left(A_{u}D_{v}-A_{v}D_{u}\right),\label{eq:BC_commutation_relation}
\end{equation}
\begin{equation}
A_{u}C_{v}=\frac{1}{b\left(u-v\right)}C_{v}A_{u}-\frac{c\left(u-v\right)}{b\left(u-v\right)}C_{u}A_{v},
\end{equation}
\begin{equation}
D_{u}C_{v}=\frac{1}{b\left(v-u\right)}C_{v}D_{u}-\frac{c\left(v-u\right)}{b\left(v-u\right)}C_{u}D_{v},
\end{equation}

\begin{equation}
\left[A_{u},D_{v}\right]=\frac{c\left(u-v\right)}{b\left(u-v\right)}\left(C_{v}B_{u}-C_{u}B_{v}\right),\label{eq:AD_commutation_relation}
\end{equation}
where the subscript for $u$ and $v$ was introduced as a shorthand
for the argument, e.g. $A_{u}\equiv A\left(u\right)$.

Within the algebraic approach the transfer matrix is given by the
trace of the transition matrix as 
\begin{equation}
\tau\left(u\right)=\mathrm{Tr}T\left(u\right)=A\left(u\right)+D\left(u\right).\label{eq:Tau_def}
\end{equation}
This operator gives a family of commuting matrices, $\left[\tau\left(u\right),\tau\left(v\right)\right]=0$
for all pairs of $u$ and $v$, which contain all the conserved quantities
of the model in Eq.~(\ref{eq:H_XXZ}) including the Hamiltonian
itself. Thus, if $\left|\mathbf{u}\right\rangle $ is an eigenstate
of $\tau\left(u\right)$, then it is also an eigenstate of the Hamiltonian.
Therefore, the eigenstate equation can be written down as 
\begin{equation}
\tau\left(u\right)\left|\mathbf{u}\right\rangle =\mathcal{T}_{u}\left|\mathbf{u}\right\rangle ,\label{eq:TAUmatrix_eigenvalue_eq}
\end{equation}
where $\mathcal{T}_{u}$ is a scalar quantity--the corresponding
eigenvalue of the transition matrix. 

The diagonalisation problem in Eq.~(\ref{eq:TAUmatrix_eigenvalue_eq})
can be solved using the commutation relation in Eqs.~(\ref{eq:BC_commutation_relation}-\ref{eq:AD_commutation_relation}).
The results of acting with the $A_{u}$ and $D_{u}$ operators on
the state $\left|\mathbf{u}\right\rangle $ in Eq.~(\ref{eq:ABA_state})
are obtained by commuting them from left to right through the product
of $C\left(u_{j}\right)$ operators,
\begin{align}
A_{u}\prod_{j=1}^{M}C\left(u_{j}\right)\left|\Downarrow\right\rangle  & =a_{u}\prod_{j=1}^{M}\frac{1}{b_{uj}}C\left(u_{j}\right)\left|\Downarrow\right\rangle -\sum_{j=1}^{M}a_{j}\frac{c_{uj}}{b_{uj}}C\left(u\right)\prod_{l=1\neq j}^{M}\frac{1}{b_{jl}}C\left(u_{l}\right)\left|\Downarrow\right\rangle ,\label{eq:A_commutation}\\
D_{u}\prod_{j=1}^{M}C\left(u_{j}\right)\left|\Downarrow\right\rangle  & =d_{u}\prod_{j=1}^{M}\frac{1}{b_{ju}}C\left(u_{j}\right)\left|\Downarrow\right\rangle +\sum_{j=1}^{M}d_{j}\frac{c_{uj}}{b_{uj}}C\left(u\right)\prod_{l=1\neq j}^{M}\frac{1}{b_{lj}}C\left(u_{l}\right)\left|\Downarrow\right\rangle ,\label{eq:D_commutation}
\end{align}
where the vacuum eigenvalues of the operators $A_{u}\left|\Downarrow\right\rangle =a_{u}\left|\Downarrow\right\rangle $
and $D_{u}\left|\Downarrow\right\rangle =d_{u}\left|\Downarrow\right\rangle $
are obtained explicitly using of the construction in Eqs.~(\ref{eq:Lmatrix_XXZ},\ref{eq:Tmatrix_Lax_construction})
as 
\begin{equation}
a_{u}=\frac{\cosh^{N}\left(u-\eta\right)}{\cosh^{N}\left(u+\eta\right)}\quad\mathrm{and}\quad d_{u}=1,\label{eq:ad_vacuum}
\end{equation}
and shorthand notations with the subscripts were introduced as $a_{j}\equiv a\left(u_{j}\right)$,
$b_{jl}\equiv b\left(u_{j}-u_{l}\right)$, and $b_{ju}\equiv b\left(u_{j}-u\right)$.

Since the right-hand sides of Eqs.~(\ref{eq:A_commutation},\ref{eq:D_commutation})
contain terms that are not proportional to the original state multiplied
by a scalar, an arbitrary algebraic Bethe state with arbitrary set
of parameters $u_{j}$ is not an eigenstate of the transfer matrix
$\tau\left(u\right)$. However, the sum of the two second terms in
Eqs.~(\ref{eq:A_commutation},\ref{eq:D_commutation}), appearing
in definition of the transition matrix in Eq.~(\ref{eq:Tau_def}),
can be made zero by selecting particular sets of $u_{j}$ that satisfy
the following set of $M$ equations, 
\begin{equation}
\frac{a_{j}}{d_{j}}=\prod_{l=1\neq j}^{M}\frac{b_{jl}}{b_{lj}}.
\end{equation}
Substitution of the expressions for $a_{j}$ and $d_{j}$ from Eq.~(\ref{eq:ad_vacuum})
and for $b_{jl}$ from Eq.~(\ref{eq:bc_XXZ_def}) gives the set
of equations for the diagonalisation problem of the transition matrix
as
\begin{equation}
\frac{\cosh^{N}\left(u_{j}-\eta\right)}{\cosh^{N}\left(u_{j}+\eta\right)}=\prod_{l=1\neq j}^{M}\frac{\sinh\left(u_{j}-u_{l}-2\eta\right)}{\sinh\left(u_{j}-u_{l}+2\eta\right)}\label{eq:ABA_equations}
\end{equation}
and its corresponding eigenvalue as 
\begin{equation}
\mathcal{T}_{u}=\frac{\cosh^{N}\left(u-\eta\right)}{\cosh^{N}\left(u+\eta\right)}\prod_{j=1}^{M}\frac{\sinh\left(u-u_{j}+2\eta\right)}{\sinh\left(u-u_{j}\right)}+\prod_{j=1}^{M}\frac{\sinh\left(u_{j}-u+2\eta\right)}{\sinh\left(u_{j}-u\right)}.\label{eq:Transition_matrix_eigenvalue}
\end{equation}

The set of equations (\ref{eq:ABA_equations}) are the Bethe equations
for the XXZ model in Eq.~(\ref{eq:H_XXZ}). Noting that the spin
momenta in the set of equations\ (\ref{eq:ABA_equations}) are written
in Orbach parametrisation \citep{Gaudin_book}, their mapping back
to the coordinate representation is given by 

\begin{align}
u_{j} & =\frac{1}{2}\ln\left(\frac{1-e^{iq_{j}-2\eta}}{1-e^{-iq_{j}-2\eta}}\right)-\frac{iq_{j}}{2}\label{eq:ABA_coordinate_mapping_uj}\\
\eta & =\frac{1}{2}\mathrm{acosh}\Delta.\label{eq:ABA_coordinate_mapping_eta}
\end{align}
Substitution of this mapping into Eq.~(\ref{eq:ABA_equations})
gives the Bethe ansatz equations\ (\ref{eq:LiebWu_spin_SM}) with
the scattering phases for the XXZ model in Eq.~(\ref{eq:2body_scattering_phase_XXZ}).

\subsection{Scalar product of Bethe states}

A straightforward example of the advantage that the algebraic representation
of Bethe ansatz provides over the coordinate representation is calculation
of the scalar product between two Bethe states $\left\langle \mathbf{v}\right|$
and $\left|\mathbf{u}\right\rangle $ given in the algebraic representation
by Eq.~(\ref{eq:ABA_state}). It can be evaluated directly with
the help of the commutation relations in Eqs.~(\ref{eq:BC_commutation_relation}-\ref{eq:AD_commutation_relation}).

The multiplication of the bra and ket states is the vacuum expectation
value of a product of the Bethe ansatz operators $B\left(v_{j}\right)$
and $C\left(u_{j}\right)$, in which all the $B\left(v_{j}\right)$
operators appear to the left from all the $C\left(u_{j}\right)$ operators.
Under such an expectation value, each $B\left(v_{j}\right)$ operator
can be commuted all the way from left to right through the product
of $C\left(u_{j}\right)$ operators using the commutation relation
in Eq.~(\ref{eq:BC_commutation_relation}), which generates all
the $A$ and $D$ operators with arguments that all possible values
of $u_{j}$ and $v_{j}$. They, in turn, also have to be commuted
to the right through the remaining product of $C\left(u_{j}\right)$
operators. Finally, the $B\left(v_{j}\right)$ operators acting upon
the vacuum state give $0$ and the products of $A$ and $D$ operators
give products of their vacuum eigenvalues $a\left(u_{j}\right)$ and
$d\left(v_{j}\right)$ given in Eq.~(\ref{eq:ad_vacuum}). When
at least one set of the parameters, say $u_{j}$, is a solution of
the Bethe equations (\ref{eq:ABA_equations}) the result of all the
commutations can be written in a compact form as a determinant of
an $M\times M$ matrix--the so-called Slavnov's formula \citep{Slavnov89},

\begin{equation}
\left\langle \mathbf{v}|\mathbf{u}\right\rangle =\frac{\prod_{ij}^{M}\sinh\left(v_{j}-u_{j}\right)}{\prod_{i<j}^{M}\sinh\left(v_{i}-v_{j}\right)\prod_{i<j}^{M}\sinh\left(u_{i}-u_{j}\right)}\det\hat{C},\label{eq:Scalar_product_ABA}
\end{equation}
where the matrix elements of the $M\times M$ matrix $\hat{C}$ are
$C_{ab}=\partial_{u_{a}}\mathcal{T}\left(v_{b}\right)$. Under substitution
of the eigenvalue of the transition matrix $\mathcal{T}\left(u\right)$
from Eq.~(\ref{eq:Transition_matrix_eigenvalue}), these matrix
elements read in explicit form as 
\begin{equation}
C_{ab}=\frac{\cosh^{N}\left(v_{b}-\eta\right)}{\cosh^{N}\left(v_{b}+\eta\right)}\frac{\sinh\left(2\eta\right)}{\sinh^{2}\left(v_{b}-u_{a}\right)}\prod_{j=1\neq a}^{M}\frac{\sinh\left(v_{b}-u_{j}+2\eta\right)}{\sinh\left(v_{b}-u_{j}\right)}-\frac{\sinh\left(2\eta\right)}{\sinh^{2}\left(u_{a}-v_{b}\right)}\prod_{j=1\neq a}^{M}\frac{\sinh\left(u_{j}-v_{b}+2\eta\right)}{\sinh\left(u_{j}-v_{b}\right)}.
\end{equation}

\subsection{Normalisation factor of Bethe state}

The normalisation factor of the Bethe states in the algebraic form
in Eq.~(\ref{eq:ABA_state}) can be evaluated by taking the limit
of $\mathbf{v}\rightarrow\mathbf{u}$ in the scalar product in Eq.~(\ref{eq:Scalar_product_ABA})
giving \citep{Gaudin81,Korepin82}
\begin{equation}
Z^{2}=\left\langle \mathbf{u}|\mathbf{u}\right\rangle =\sinh^{M}\left(2\eta\right)\prod_{i\neq j}^{M}\frac{\sinh\left(u_{j}-u_{i}+2\eta\right)}{\sinh\left(u_{j}-u_{i}\right)}\det\hat{F}\label{eq:Normalisation_ABA}
\end{equation}
where the matrix elements of $\hat{F}$ are 

\begin{equation}
F_{ab}=\begin{cases}
-N\frac{\sinh\left(2\eta\right)}{\cosh\left(u_{a}+\eta\right)\cosh\left(u_{a}-\eta\right)}-\sum_{j=1\neq a}^{M}\frac{\sinh\left(4\eta\right)}{\sinh\left(u_{a}-u_{j}-2\eta\right)\sinh\left(u_{a}-u_{j}+2\eta\right)}, & a=b,\\
\frac{\sinh\left(4\eta\right)}{\sinh\left(u_{b}-u_{a}-2\eta\right)\sinh\left(u_{b}-u_{a}+2\eta\right)}, & a\neq b.
\end{cases}
\end{equation}

Let us map this result back into the coordinate representation by
substituting Eqs.~(\ref{eq:ABA_coordinate_mapping_uj},\ref{eq:ABA_coordinate_mapping_eta})
for the Heisenberg model $\Delta=1$, which is a part of the original
problem for the spin part of the wave function in this work. For $\Delta=1$
the interaction parameter in the algebraic representation in Eq.~(\ref{eq:ABA_coordinate_mapping_eta})
is $\eta=0$ making the mapping for spin momenta in Eq.~(\ref{eq:ABA_coordinate_mapping_uj})
degenerate, $u_{j}=i\pi/2$ for any $q_{j}$. Therefore, the proper
limit needs to be taken by expanding Eq.~(\ref{eq:ABA_coordinate_mapping_uj})
in Taylor series upto the linear order in $\eta$, 
\begin{equation}
u_{j}=\frac{i\pi}{2}+i\eta\cot\frac{q_{j}}{2},\label{eq:uj_expansion}
\end{equation}
substituting this expansion in the normalisation factor in Eq.~(\ref{eq:Normalisation_ABA}),
and taking the limit of $\eta\rightarrow0$ for the whole expression.
Implementation of this procedure, 
\begin{equation}
\lim_{\eta\rightarrow0}\left.Z^{2}\right|_{u_{j}=\frac{i\pi}{2}+i\eta\cot\frac{q_{j}}{2}},
\end{equation}
gives 
\begin{equation}
Z^{2}=\left(-4\right)^{M}\prod_{i\neq j=1}^{M}\frac{\left(\cot\frac{q_{j}}{2}-\cot\frac{q_{i}}{2}-2i\right)}{\left(\cot\frac{q_{j}}{2}-\cot\frac{q_{i}}{2}\right)}\prod_{j}^{M}\sin^{2}\frac{q_{j}}{2}\det\hat{Q},\label{eq:Normalisation_coordinate}
\end{equation}
where the matrix under the determinant is 
\begin{equation}
Q_{ab}=\begin{cases}
N-\sum_{j=1\neq a}^{M}\frac{4\left(1-\cos q_{j}\right)}{\left(e^{iq_{j}}+e^{-iq_{a}}-2\right)\left(e^{iq_{a}}+e^{-iq_{j}}-2\right)} & ,a=b,\\
\frac{4\left(1-\cos k_{b}\right)}{\left(e^{iq_{b}}+e^{-iq_{a}}-2\right)\left(e^{iq_{a}}+e^{-iq_{b}}-2\right)} & ,a\neq b.
\end{cases}.\label{eq:Qmatrix}
\end{equation}

This is the determinant of the same matrix as was obtained for the
normalisation of the Bethe state in the coordinate representation
in \citep{Gaudin81}, which is quoted after Eq.~(5) in the main
paper. However, the prefactor in front of this determinant is not
$1$ but a function of spin momenta, see Eq.~(\ref{eq:Normalisation_coordinate}),
demonstrating that normalisation of the wave functions in two different
representations in Eq.~(\ref{eq:Psi_s_SM}) and in Eq.~(\ref{eq:ABA_state})
is different. Since the algebraic representation is used for the calculation
of the correlation functions in this work, Eq.~(\ref{eq:Normalisation_coordinate})
will be used as the normalisation factor for the relevant spin matrix
elements, although the final result, Eqs.~(\ref{eq:ME_spin_SM}-\ref{eq:RMb_SM})
below and Eqs.~(7-9) of the main paper, is presented in the coordinate
representation. 

\section{Spin matrix element }

Evaluation of the spin part, $\big<f\big|c_{1\uparrow}\big|0\big>_{s}$,
of the matrix element, needed for the correlation functions in the
main paper, requires the use of algebraic representation of Bethe
ansatz introduced in the previous section. Since the spin part of
the wave function does not contain any change degree of freedom, the
explicit form of this matrix elements is 
\begin{equation}
\big<f\big|c_{1\uparrow}\big|0\big>_{s}=Z_{f}^{-1}Z_{0}^{-1}\big<f\big|D_{1}S_{1}^{-}\big|0\big>_{s},\label{eq:ME_c1_SM}
\end{equation}
were $Z$ are the normalisation factor of the Bethe wave function
in the algebraic representation in Eq.~(\ref{eq:Normalisation_ABA}).
Note that the charge part of the $c_{1\uparrow}$ operator, $a_{1}$,
acts only on the charge degrees of freedom giving the result quoted
in Eq.~(6) of the main text \citep{Penc96}. 

There are two problems in calculating the spin matrix element in Eq.\ (\ref{eq:ME_c1_SM}).
One is representation of the local spin operator $S_{1}^{-}$ in terms
of the delocalised along the chain Bethe operators in Eq.~(\ref{eq:Tmatrix_def})
so the commutation relations in Eqs.~(\ref{eq:BC_commutation_relation}-\ref{eq:AD_commutation_relation})
can be used. And the other problem is dealing with the $D_{1}$ operator
that makes the spin chain shorter by one site and, therefore, the
Bethe operators in the algebraic representation for the states $\big|0\big>_{s}$
and $\big<f\big|_{s}$ in Eq.~(\ref{eq:ABA_state}) are different
according to the construction in Eq.~(\ref{eq:Tmatrix_Lax_construction})
so the commutation relations between them are not defined. We will
address these two problems one by one.

The first problem of expressing the local spin operators in terms
of Bethe operators was solve by means of Drinfeld twist \citep{Drinfeld83}.
It allows to construct a representation for the transition matrix
in which it becomes quasi-local, obtaining a relatively simple expression
for the original spin operators. For instance, for the $S_{j}^{-}$
operator, appearing in matrix element in Eq.~(\ref{eq:ME_c1_SM}),
it reads \cite{Kitanine99,*Kitanine00}
\begin{equation}
S_{j}^{-}=\tau_{\xi}^{N-j}B_{\xi}\tau_{\xi}^{j-1}\label{eq:Smj_ABA}
\end{equation}
where $\xi=-i\pi/2+\eta$. 

Note that this expression is not identical to the one presented in
\cite{Kitanine99,*Kitanine00}. The correctness of the above expression
can be check explicitly by expressing the $B_{\xi}$ and $\tau_{\xi}$
operators in terms of the local spin operators for a chain of small
length $N=3$, by means of the construction in Eq.~(\ref{eq:Tmatrix_Lax_construction}).
As a result, it can be checked explicitly that the transfer matrix
$\tau_{\xi}$ shifts the chain by one site to the left and the Bethe
operator $B_{\xi}$ removes the last spin in the chain. The discrepancy
between Eq.~(\ref{eq:Smj_ABA}) and \cite{Kitanine99,*Kitanine00}
makes no difference for the pure Heisenberg model since the same periodic
boundary condition for the bra and the ket states ensures a change
only in the irrelevant phase factor in the matrix elements there.
But using one or the other expression strongly affects the correlation
function of the Hubbard model, in which the relevant ladder operators
simultaneously change the length of the spin chain, leading to different
amplitudes in the resulting matrix elements--not just different phase
factors.

Still in the coordinate representation, the spin chain can be shifted
$N-1$ times to the right. Under such a shift, the matrix element
in Eq.~(\ref{eq:ME_c1_SM}) acquires only an irrelevant phase factor, 

\begin{equation}
\big<f\big|D_{1}S_{1}^{-}\big|0\big>_{s}=e^{i\left(P_{s}^{0}-P_{s}^{f}\right)\left(N-1\right)}\left\langle f|D_{N}S_{N}^{-}|0\right\rangle .\label{eq:ME_SmN}
\end{equation}
However, it will help to avoid the need to commute the operators $\tau_{\xi}$
and $B_{\xi}$ explicitly, simplifying the algebra in the calculation
below. 

Now, the whole the whole spin matrix element can be written in the
algebraic representation by substituting the expression in Eq.~(\ref{eq:Smj_ABA})
for $S_{N}^{-}$ and the Bethe wave function in Eq.~(\ref{eq:ABA_state})
for the eigenstates into the matrix element in the right-hand side
of Eq.~(\ref{eq:ME_SmN}) as 

\begin{equation}
\left\langle f|D_{N}S_{N}^{-}|0\right\rangle =\big\langle\Downarrow\big|\prod_{j=1}^{M-1}B^{N-1}\left(v_{j}\right)B_{\xi}^{N}\prod_{j=1}^{M}C^{N}\left(u_{j}\right)\big|\Downarrow\big\rangle,\label{eq:ME_SmN_ABA}
\end{equation}
where the set of $u_{j}$ and the set of $v_{j}$ are two solutions
of the Bethe equations\ (\ref{eq:ABA_equations}) and $\tau_{\xi}^{N-1}$
acting upon an eigenstate, see Eq.~(\ref{eq:TAUmatrix_eigenvalue_eq}),
gives just its eigenvalue in Eq.~(\ref{eq:Transition_matrix_eigenvalue}),
which for $u=\xi$ becomes just an irrelevant phase shift $\mathcal{T}_{\xi}^{N-1}\big|0\big\rangle=e^{iP_{s}^{0}\left(N-1\right)}\big|0\big\rangle$.
Here the $D_{N}$ operator removes the $N^{\mathrm{th}}$ spin from
the chain, reducing its length by one. Therefore, the Bethe ansatz
operators before the position of this operators are constructed, see
Eq.~(\ref{eq:Tmatrix_Lax_construction}), out of the spin operators
for the $1^{\mathrm{st}}$ to $N^{\mathrm{th}}$ site, which is marked
by the superscript $N$, \emph{e.g.} $C^{N}\left(u_{j}\right)$. The
Bethe ansatz operators after the position of the $D_{N}$ operator
are constructed out of the spin operators for the $1^{\mathrm{st}}$
to $N-1^{\mathrm{st}}$ site, which is marked by the superscript $N-1$,
\emph{e.g.} $B^{N-1}\left(v_{j}\right)$.

The operators with the superscript $N$ in Eq.~(\ref{eq:ME_SmN_ABA})
obey the same commutation relations in Eqs.~(\ref{eq:BC_commutation_relation}-\ref{eq:AD_commutation_relation}).
Thus, commuting $B^{N}$ through a product of $C^{N}$, a procedure
similar to what was used in deriving Eqs.~(\ref{eq:A_commutation}-\ref{eq:D_commutation}),
gives

\begin{equation}
B^{N}\left(\xi\right)\prod_{j=1}^{M}C^{N}\left(u_{j}\right)\left|\Downarrow\right\rangle =\sum_{x=1}^{M}a_{x}c_{x\xi}\prod_{i=1\neq x}^{M+1}\frac{1}{b_{xi}}\sum_{y=1\neq x}^{M+1}d_{y}c_{\xi y}\prod_{j=1\neq x,y}^{M+1}\frac{1}{b_{jy}}\prod_{j=1\neq x,y}^{M+1}C^{N}\left(u_{j}\right)\left|\Downarrow\right\rangle ,\label{eq:B_commutation}
\end{equation}
where the notation of $u_{M+1}\equiv\xi$ was introduced and all the
vacuum eigenvalues and the matrix element of the $R$-matrix are given
only for $u_{j}$, \emph{e.g.} $d_{x}=d\left(u_{x}\right)$ or $c_{x\xi}=c\left(u_{x}-\xi\right)$.
After substitution of this result back into Eq.~(\ref{eq:ME_SmN_ABA}),
the remaining structure in terms of Bethe ansatz operators is that
of a scalar product, which can, in principle, be evaluated in terms
of a determinant expression, see Eq.~(\ref{eq:Scalar_product_ABA}).
However, before the formula in Eq.~(\ref{eq:Scalar_product_ABA})
can be used both the bra and ket states have to be expressed in terms
of the Bethe ansatz operators for a spin chain of the same length
satisfying the same commutation relations.

We resolve this problem by expressing the Bethe ansatz operators for
the longer chain in terms of the Bethe ansatz operators from the shorter
chain and the local spin operator for the $N^{\mathrm{th}}$ spin.
The last ($N^{\mathrm{th}}$) spin in the product in the construction
in Eq.~(\ref{eq:Tmatrix_Lax_construction}) can be singled out giving 

\begin{equation}
\left(\begin{array}{cc}
A_{u}^{N} & B_{u}^{N}\\
C_{u}^{N} & D_{u}^{N}
\end{array}\right)=\left(\begin{array}{cc}
A_{u}^{N-1} & B_{u}^{N-1}\\
C_{u}^{N-1} & D_{u}^{N-1}
\end{array}\right)\left(\begin{array}{cc}
\frac{\cosh\left(u+\eta2S_{N}^{z}\right)}{\cosh\left(u+\eta\right)} & -i\frac{\sinh2\eta S_{N}^{-}}{\cosh\left(u+\eta\right)}\\
-i\frac{\sinh2\eta S_{N}^{+}}{\cosh\left(u+\eta\right)} & \frac{\cosh\left(u-\eta2S_{N}^{z}\right)}{\cosh\left(u+\eta\right)}
\end{array}\right),
\end{equation}
where the transition matrix for the shorter chain of $N-1$ spins,
$T^{N-1}\left(u\right)=\prod_{j=1}^{N-1}L_{j}\left(u\right)$, gives
the corresponding Bethe operators in the right-hand side. The bottom-left
matrix elements in both sides of this equation gives representation
of the $C_{u}^{N}$ operator needed for calculating the element in
Eq.~(\ref{eq:ME_SmN_ABA}) through Bethe operators of the shorter
chain,

\begin{equation}
C^{N}=\frac{\cosh\left(u+\eta2S_{N}^{z}\right)}{\cosh\left(u+\eta\right)}C^{N-1}-i\frac{\sinh2\eta S_{N}^{+}}{\cosh\left(u+\eta\right)}D^{N-1}.\label{eq:CN_CNm1}
\end{equation}

Substituting the result of the commutation in Eq.~(\ref{eq:B_commutation})
and this expression above into Eq.~(\ref{eq:ME_SmN_ABA}), we obtain
a vacuum expectation value with respect to the ferromagnetic states
of the chain of $N$ spins, $\left|\Downarrow\right\rangle ^{N}=\prod_{j=1}^{N}\left|\downarrow\right\rangle _{j}$.
In this expression, the contribution of all terms containing $S_{N}^{+}$
at least in the first power is zero since the product of $B^{N-1}$
contains no $S_{N}^{-}$ operators so that such terms has a factor
\begin{equation}
\big<\downarrow\big|\left[\frac{\sinh2\eta S_{N}^{+}}{\cosh\left(u+\eta\right)}\right]^{k}\big|\downarrow\big>_{N}=0
\end{equation}
for $k>0$. The other terms containing only $S_{N}^{z}$ operators
give a non-zero contribution,

\begin{equation}
\big<\downarrow\big|\left[\frac{\cosh\left(u+\eta2S_{N}^{z}\right)}{\cosh\left(u+\eta\right)}\right]^{k}\big|\downarrow\big>_{N}=\frac{\cosh^{k}\left(u-\eta\right)}{\cosh^{k}\left(u+\eta\right)}.
\end{equation}
The sum of all such uch non-zero terms gives the matrix element in
Eq.~(\ref{eq:ME_SmN_ABA}) as 

\begin{align}
\left\langle f|D_{N}S_{N}^{-}|0\right\rangle  & =\sum_{x=1}^{N}a_{x}\frac{c_{x\xi}}{b_{x\xi}}\prod_{i=1\neq x}^{N}\frac{1}{b_{xi}}\sum_{y=1\neq x}^{N}\frac{c_{\xi y}}{b_{\xi y}}\frac{\cosh\left(\xi-\eta\right)}{\cosh\left(\xi+\eta\right)}\prod_{j=1\neq x,y}^{N}\frac{1}{b_{jy}}\frac{\cosh\left(u_{j}-\eta\right)}{\cosh\left(u_{j}+\eta\right)}\left\langle \mathbf{v}|u_{x-1},u_{x+1},u_{y-1},u_{y+1},\xi\right\rangle \label{eq:ME_intermediate_result_line1}\\
 & +\sum_{x=1}^{N}a_{x}\frac{c_{x\xi}}{b_{x\xi}}\prod_{i=1\neq x}^{N}\frac{1}{b_{xi}}\prod_{j=1\neq x}^{N}\frac{1}{b_{j\xi}}\frac{\cosh\left(u_{j}-\eta\right)}{\cosh\left(u_{j}+\eta\right)}\left\langle \boldsymbol{v}|u_{x-1},u_{x+1}\right\rangle .\\
= & \prod_{j=1}^{N}\frac{\cosh\left(u_{j}-\eta\right)}{\cosh\left(u_{j}+\eta\right)}\sum_{x=1}^{N}a_{x}\frac{c_{x\xi}}{b_{x\xi}}\prod_{i=1\neq x}^{N}\frac{1}{b_{xi}}\frac{\cosh\left(u_{x}+\eta\right)}{\cosh\left(u_{x}-\eta\right)}\prod_{j=1\neq x}^{N}\frac{1}{b_{j\xi}}\left\langle \boldsymbol{v}|u_{x-1},u_{x+1}\right\rangle ,\label{eq:ME_intermediate_result_line3}
\end{align}
where the factor $\cosh\left(\xi-\eta\right)/\cosh\left(\xi+\eta\right)=0$
makes the line\ (\ref{eq:ME_intermediate_result_line1}) zero. Here
the scalar factors can be taken out of the vacuum expectation value,
leaving only a product of $M-1$ operators $B^{N-1}$ and $C^{N-1}$
under it, in which all the $B^{N-1}$ operators occur to the left
of all the $C^{N-1}$ operators, $\left\langle \boldsymbol{v}|u_{x-1},u_{x+1}\right\rangle $.
The latter is evaluated using the determinant formula for scalar product
in Eq.~(\ref{eq:Scalar_product_ABA}), where the elements of the
matrix under the determinant are $C_{ab}=\partial_{v_{a}}\mathcal{T}\left(u_{b}\right)$
since $v_{1},\dots,v_{M-1}$ is a solution of Bethe equations for
the shorter chain but the set of $M-1$ parameters $u_{1},\dots,u_{x-1},u_{x+1},\dots,u_{M}$
is not a solution of any Bethe equation.

The remaining sum over determinants of $M-1\times M-1$ matrices in
Eq.~(\ref{eq:ME_intermediate_result_line3}) can be expressed as
a single determinant of an $M\times M$ matrix since the sum has the
form of a Laplace expansion of a matrix through the sum of minors
over one of its columns,
\begin{equation}
\det\hat{T}=\sum_{x=1}^{M}T_{Mx}\left(-1\right)^{x+M}\mathrm{minor}_{Mx}.
\end{equation}
where $T_{Mx}$ is the entry of $M^{\mathrm{th}}$ row and $x^{\mathrm{th}}$
column of the matrix $\hat{T}$ and $\mathrm{minor}_{Mx}$ is the
determinant of the submatrix obtained by removing the $M^{\mathrm{th}}$
row and $x^{\mathrm{th}}$ column of $\hat{T}$. Applying this formula
Eq.~(\ref{eq:ME_intermediate_result_line3}) in reverse, and rearranging
various factors for compactness, we obtain the following expression
for the matrix element
\begin{equation}
\left\langle f|D_{N}S_{N}^{-}|0\right\rangle =\frac{\prod_{i,j}^{M-1,M}\sinh\left(u_{j}-v_{i}\right)}{\prod_{j<i}^{M-1}\sinh\left(v_{j}-v_{i}\right)\prod_{j<i}^{M}\sinh\left(u_{j}-u_{i}\right)}\det\hat{R},\label{eq:ME_det_result_ABA}
\end{equation}
where the elements of the $M\times M$ matrix $\hat{R}$ are 
\begin{equation}
R_{ab}=\frac{\sinh\left(2\eta\right)}{\sinh^{2}\left(u_{b}-v_{a}\right)}\Bigg[\frac{\cosh\left(u_{b}-\eta\right)^{N-1}}{\cosh\left(u_{b}+\eta\right)^{N-1}}\prod_{j=1\neq a}^{M-1}\frac{\sinh\left(u_{b}-v_{j}+2\eta\right)}{\sinh\left(u_{b}-v_{j}\right)}-\prod_{j=1\neq a}^{M-1}\frac{\sinh\left(v_{j}-u_{b}+2\eta\right)}{\sinh\left(v_{j}-u_{b}\right)}\Bigg],
\end{equation}
for $a<M$ and 
\begin{equation}
R_{Mb}=\frac{\prod_{i=1}^{M}\sinh\left(u_{b}-u_{i}-2\eta\right)}{\cosh\left(u_{b}-\eta\right)\prod_{i}^{M-1}\sinh\left(u_{b}-v_{i}\right)}
\end{equation}
for $a=M$. Calculation of the the same matrix element in the coordinate
representation for a small number of spin excitations $M=1,2,3$,
using the Bethe wave function in Eqs.~(\ref{eq:Psi_s_SM},\ref{eq:2body_scattering_phase_XXZ})
and evaluating the $M$-fold integrals numerically, give the same
result as Eq.~(\ref{eq:ME_det_result_ABA}). 

Finally, Eq.~(\ref{eq:ME_det_result_ABA}) needs to be substituted
back into Eqs.~(\ref{eq:ME_SmN},\ref{eq:ME_c1_SM}) and the Heisenberg
limit of $\Delta=1$ has to be taken. The latter is done by substituting
the Taylor expansion for the $u_{j}$ and $v_{j}$ in small $\eta$
from Eq.~(\ref{eq:uj_expansion}) and taking the limit of the resulting
expression 

\begin{equation}
\left\langle f|c_{j\uparrow}|0\right\rangle _{s}=\lim_{\eta\rightarrow0}\frac{\left\langle f|D_{N}S_{N}^{-}|0\right\rangle }{Z_{0}Z_{f}},
\end{equation}
giving

\begin{equation}
\left\langle f|c_{j\uparrow}|0\right\rangle _{s}=\frac{1}{\sqrt{\det\hat{Q}_{0}\det\hat{Q}_{f}}}\frac{\prod_{i,j}^{M-1,M}\left(e^{iq_{i}^{f}}+e^{-iq_{j}^{0}}-2\right)}{\prod_{i\neq j}^{M-1}\sqrt{e^{iq_{i}^{f}}+e^{-iq_{j}^{f}}-2}\prod_{i\neq j}^{M}\sqrt{e^{iq_{i}^{0}}+e^{-iq_{j}^{0}}-2}}\det\hat{R},\label{eq:ME_spin_SM}
\end{equation}
where the elements of the $M\times M$ matrix $\hat{R}$ are 
\begin{align}
R_{ab} & =\frac{e^{iq_{b}^{0}\left(N-1\right)}\prod_{j\neq a}^{M-1}\left(-\frac{e^{iq_{j}^{f}+iq_{b}^{0}}+1-2e^{iq_{j}^{f}}}{e^{iq_{j}^{f}+iq_{b}^{0}}+1-2e^{iq_{b}^{0}}}\right)-1}{\left(e^{-iq_{a}^{f}}-e^{-iq_{b}^{0}}\right)\left(e^{iq_{a}^{f}}+e^{-iq_{b}^{0}}-2\right)},\label{eq:Rab_SM}\\
R_{Mb} & =\frac{e^{ik_{b}^{0}}\prod_{i\neq b}^{M}\left(e^{iq_{i}^{0}}+e^{-iq_{b}^{0}}-2\right)}{\prod_{j}^{M-1}\left(e^{iq_{i}^{f}}+e^{-iq_{b}^{0}}-2\right)},\label{eq:RMb_SM}
\end{align}
and the matrices $\hat{Q}_{0}$ and $\hat{Q}_{f}$ are given by Eq.~(\ref{eq:Qmatrix})
for the spin momenta of the initial $0$ and the final state $f$
respectively. Repetition of the same calculation for $\big\langle f\big|c_{1j}^{\dagger}\big|0\big\rangle _{s}$
gives the same expression as in Eqs.~(\ref{eq:ME_spin_SM}-\ref{eq:RMb_SM}),
in which the spin momenta are swapped, $\mathbf{q}^{0}\leftrightarrow\mathbf{q}^{f}$,
and the spin and particle quantum numbers are increased by one, $N\rightarrow N+1$
and $M\rightarrow M+1$.

The result in Eqs.~(\ref{eq:ME_spin_SM}-\ref{eq:RMb_SM}) is presented
in Eq.~(7-9) of the main paper. 

\bibliographystyle{apsrev4-2}
\bibliography{citations}

\begin{figure*}[p]
\centering

\includegraphics[width=0.8\columnwidth]{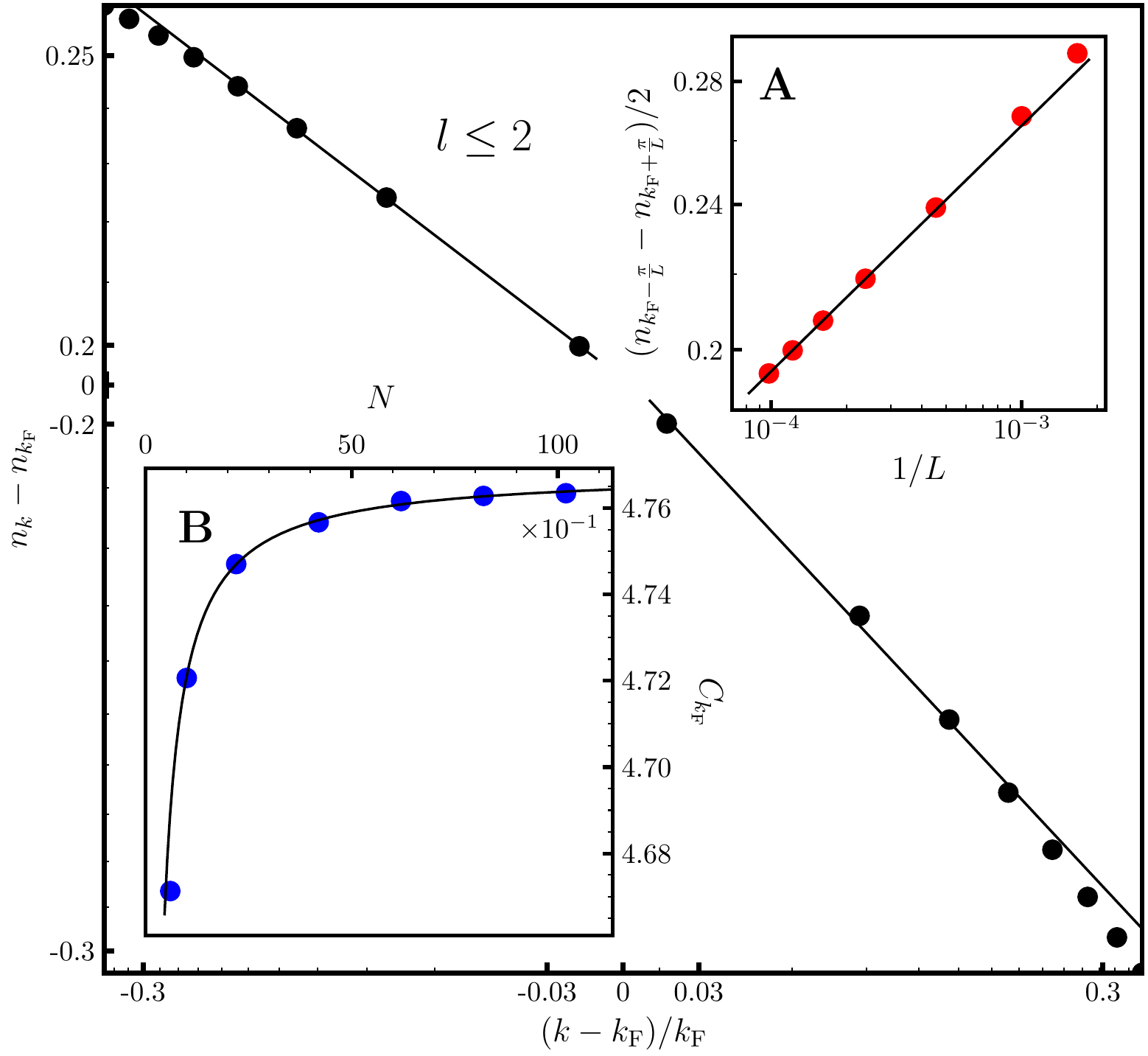}

\caption{Momentum distribution function $n_{k}$ around the $k_{\mathrm{F}}$
point on the log-log scale for $N=80$ particles, where the three
leading levels of the hierarchy of modes $l\protect\leq2$ were taken
into account in the sum in Eq.~(11) of the main text. The dashed
solid lines are power-law functions for $k<k_{\mathrm{F}}$ and for
$k>k_{\mathrm{F}}$ giving the exponent as $a_{k_{\mathrm{F}}}=0.124\pm0.020$,
where the value is the average of the two and the error bars is the
difference. Inset A: Finite size cutoff for $n_{k}$ as the function
of inverse system size $1/L$ on the log-log scale at the $k_{\mathrm{F}}$
point. The solid solid line is a power-law fit giving $a_{k_{\mathrm{F}}}=0.133884$,
within the accuracy of the fitting $n_{k}$ directly. Inset B: The
value of $n_{k}$ at the $k_{\mathrm{F}}$ point as a function of
the particle number $N$. The solid line is a finite size a fit with
finite size corrections, $n_{k_{\mathrm{F}}}=C_{k_{\mathrm{F}}}+b/N$
giving $C_{k_{\mathrm{F}}}=0.477$.}
\end{figure*}

\begin{figure*}[p]
\centering

\includegraphics[width=0.8\columnwidth]{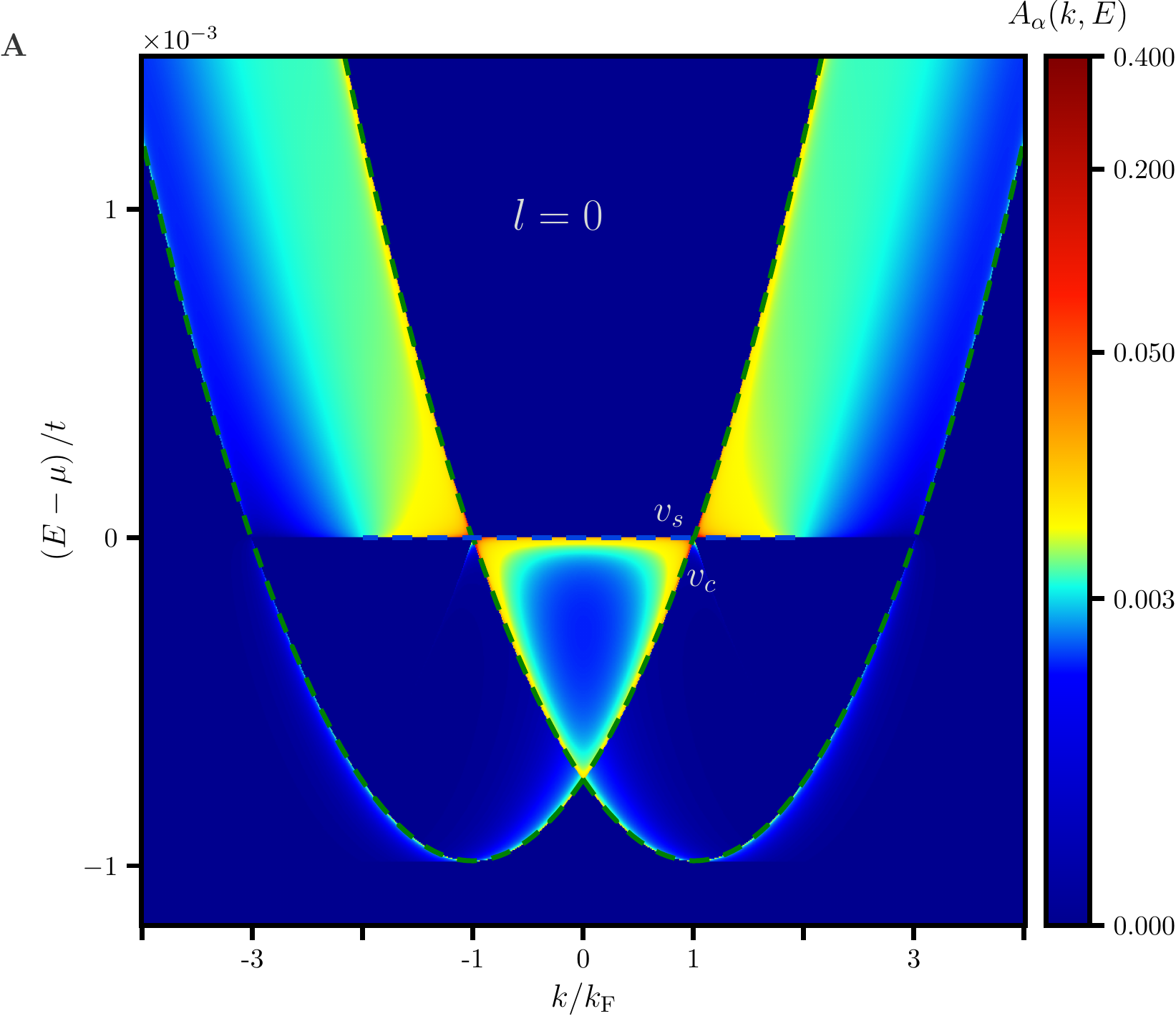}

\medskip{}

\includegraphics[width=0.8\columnwidth]{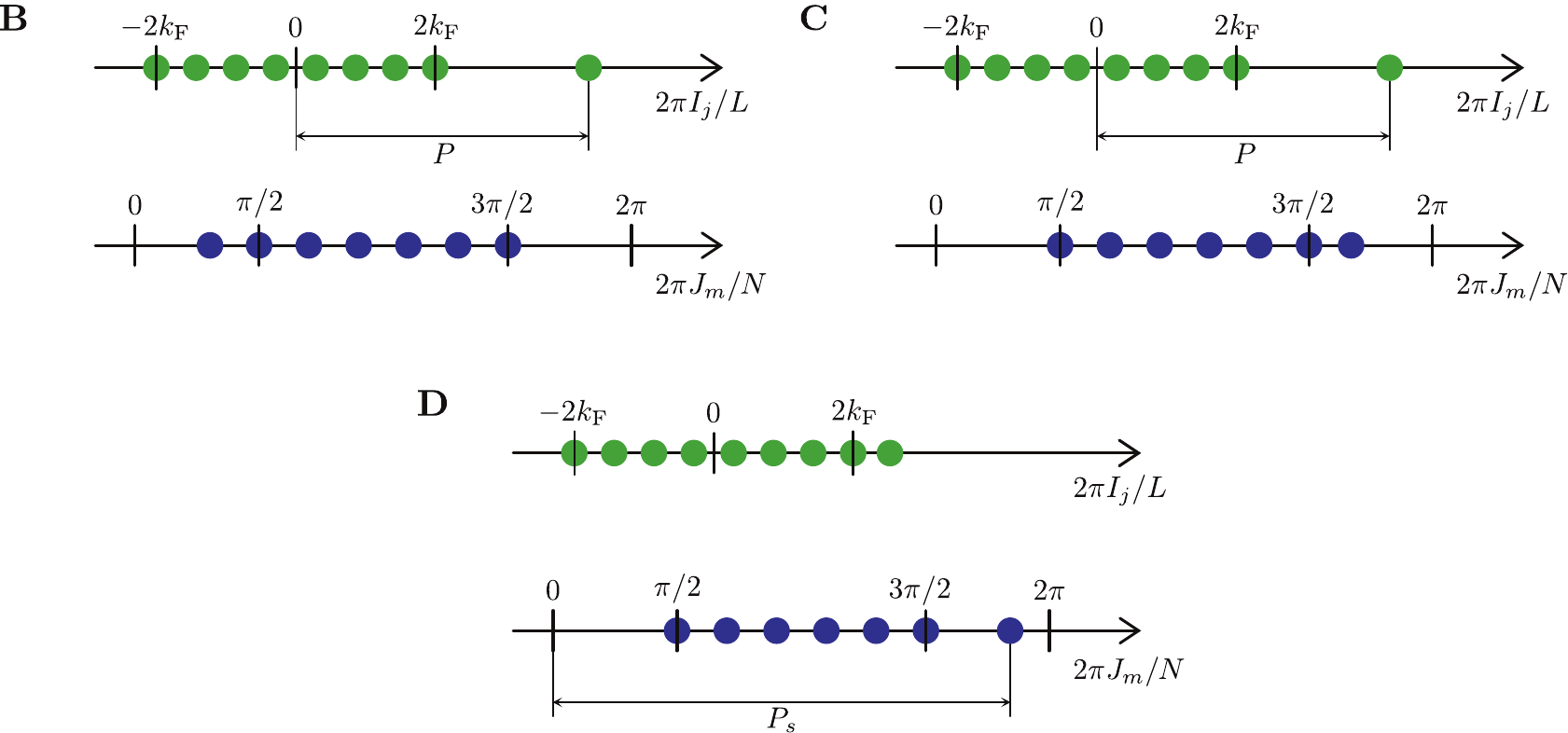}

\caption{(\textbf{A}) Spectral function $A_{\alpha}\left(k,E\right)$ of the
Hubbard model in Eq.~(1) of the main text evaluated in the $U/t=\infty$
limit using Eqs.~(6-10) of the main text for $N=500$ particles,
where only the leading level $l=0$ of the hierarchy of modes was
taken into account in the sum in Eq.~(10) of the main text. The
nonlinear holon dispersions for the states in (\textbf{B}) and (\textbf{C})
are the two green dashed lines. The nonlinear spinon dispersion for
the states in (\textbf{D}) is the blue dashed lines. Around the $\pm k_{\mathrm{F}}$
points the slopes of these dispersions are the spinon $v_{s}=0$ (labelled
as $v_{s}$) and the holon $v_{c}=2v_{\mathrm{F}}$ (labelled as $v_{c}$)
velocity, phenomenological parameters of the linear TLL model; $v_{\mathrm{F}}$
is the Fermi velocity of the free particles. (\textbf{B}) and (\textbf{C})
Two sets of integer numbers, $I_{j}$ for charge and $J_{m}$ for
spin degrees of freedom, defining the Lieb-Wu for the pure holon excitations
in the particle sector. (\textbf{D}) Two sets of integer numbers,
$I_{j}$ for charge and $J_{m}$ for spin degrees of freedom, defining
the Lieb-Wu for the pure spinon excitations in the particle sector.
The excitations in the hole sector are constructed analogously.}
\end{figure*}